\documentclass[prd,aps,11pt,
nofootinbib,tightenlines,
superscriptaddress,preprintnumbers,eqsecnum]{revtex4}

\usepackage{graphicx}
\usepackage{amsmath}
\usepackage{amssymb}

\newcommand{\be}{\begin{equation}}
\newcommand{\ee}{\end{equation}}
\newcommand{\bea}{\begin{eqnarray}}
\newcommand{\eea}{\end{eqnarray}}
\newcommand{\nn}{\nonumber}
\newcommand{\0}{\over }

\newcommand{\2}{{1\over2}}

\newcommand{\6}{\partial }
\newcommand{\9}{}

\newcommand{\tr}{\,{\rm tr}\,}

\newcommand{\asinh}{{\rm asinh}\,}

\begin{document}

\title{
Collective 
modes and 
instabilities in \\
anisotropically expanding ultarelativistic plasmas}

\preprint{TUW-09-20}

\author{Anton Rebhan}
\author{Dominik Steineder}
\affiliation{Institut f\"ur Theoretische Physik, Technische Universit\"at Wien,
        Wiedner Hauptstrasse 8-10, A-1040 Vienna, Austria}

\date{\today}


\begin{abstract}
The time evolution of collective modes in an expanding ultarelativistic and (effectively) Abelian plasma is studied in the hard-loop approximation semi-analytically by means of integro-differential equations. A previous treatment is generalized to arbitrary orientation of wave vectors with respect to the direction of anisotropy and thus to a fully
3+1 dimensional situation. Moreover, initial fluctuations are allowed in both gauge fields and currents, which is necessary in the case of (stable) longitudinal modes. For unstable (Weibel) modes, this generalization of initial conditions reduces drastically the lower bound on the delay in the onset of growth that was found previously by considering only collective gauge fields as seeds. 
This makes it appear much more likely that non-Abelian plasma instabilities seeded by small initial rapidity fluctuations could play an important role in the early stage of heavy-ion collisions at LHC energies.
\end{abstract}

\maketitle

\section{Introduction}

Fits of  hydrodynamical models to the experimental results
at the Relativistic Heavy Ion Collider (RHIC)
\cite{Teaney:2003kp,Romatschke:2007mq,Song:2007ux} 
are often interpreted as an indication of an extremely fast thermalization, or at least isotropization, 
of the quark-gluon plasma that is assumed to have
formed with initial temperatures
significantly above the deconfinement temperature.
The inferred thermalization time of $\lesssim 1$ fm/c
is so short that is hard to
understand from a perturbative framework such as the (original)
bottom-up thermalization scenario \cite{Wong:1996va,Wong:1997dv,Baier:2000sb}.
Together with the low inferred value for the specific shear viscosity, 
it seems to clearly favor strong-coupling approaches, in particular those
based on gauge-gravity duality \cite{Shuryak:2008eq}.

However, as pointed out first by Ref.~\cite{Arnold:2003rq},
a weak-coupling approach has to take into account
the inevitable presence of non-Abelian (chromo-Weibel)
plasma instabilities 
\cite{Mrowczynski:1988dz,Pokrovsky:1988bm,Mrowczynski:1993qm,Randrup:2003cw},
which produce nonperturbatively large gauge fields that may
lead to important, 
qualitative modifications of bottom-up thermalization
\cite{Mueller:2005un,Bodeker:2005nv,Arnold:2005qs,Mueller:2006up,Arnold:2007cg}.
Plasma instabilities and associated turbulent phenomena
may even be responsible for a strong reduction
of the effective shear viscosity \cite{Asakawa:2006tc}.
A complete scenario, even of just the early stages of the
evolution, is still missing, but would clearly be needed to
decide, firstly, whether the thermalization of the quark-gluon
plasma at RHIC is indeed a strong-coupling phenomenon, and secondly,
what to expect if the early stage of heavy-ion collisions at the
higher energies at the Large Hadron Collider (LHC) is probing a
regime where weak-coupling approaches based on (resummed) perturbative
quantum chromodynamics (QCD) become relevant.

At sufficiently weak coupling, the collective dynamics of a
non-Abelian plasma can be described to leading order by
an effective field theory produced by integrating out the
hard modes corresponding to real (as opposed to virtual)
plasma constituents. For anisotropic plasmas, 
the resulting hard-loop effective theory 
\cite{Mrowczynski:2004kv} is a generalization of the
well-studied hard-thermal-loop effective theory 
\cite{Taylor:1990ia,Braaten:1992gm,Frenkel:1992ts}.
The corresponding effective field equations are nonlocal, but
can be made local
at the expense of introducing a continuous set of auxiliary fields
\cite{Nair:1994xs}
which arise naturally when solving gauge covariant Boltzmann-Vlasov equations
\cite{Blaizot:1994be,Blaizot:1994am,Kelly:1994dh,Blaizot:2001nr}.
In the hard loop approximation, these auxiliary fields
depend on the velocity vector of the hard particles whose
hard momentum scale is integrated out.

The instabilities in a stationary homogeneous plasma
with momentum-space anisotropy have been studied in this
approximation in 
\cite{Romatschke:2003ms,Arnold:2003rq,Romatschke:2004jh,Schenke:2006xu,Schenke:2006fz}
for the case of weak fields, where the dynamics is
effectively Abelian. In the Abelian case, instabilities
grow exponentially
until they are large enough to modify the distribution of the hard particles,
typically giving rise to fast isotropization. 
In the non-Abelian case, nonlinear self-interactions of the collective
fields may hinder this evolution, which can only be studied by a numerical
(real-time lattice) treatment. The first numerical simulations
of non-Abelian plasma instabilities have only considered
modes that are constant in the directions transverse to
the direction of momentum anisotropy \cite{Rebhan:2004ur}.
In this situation there is a short stagnation of the exponential
growth when the non-Abelian regime is entered, after which
localized Abelianization occurs together with continued exponential growth.
However, in the more generic case of fully 3+1-dimensional evolution
\cite{Arnold:2005vb,Rebhan:2005re}
such local Abelianization appears to get largely destroyed. Instead,
a turbulent cascade \cite{Arnold:2005ef,Arnold:2005qs}
is formed where the growth is reduced to a linear one
(see however \cite{Berges:2008mr} and \cite{Dumitru:2006pz}). 
This was also found
in more extensive simulations with stronger anisotropies
in \cite{Bodeker:2007fw}, although only when initial fields
were already nonperturbatively large.

In Ref.~\cite{Romatschke:2006wg} the hard-loop effective theory for
stationary anisotropic plasmas was extended to the case of
a boost-invariant, longitudinally expanding distribution of
plasma particles, which is closer to the actual
situation in the earliest stage of
heavy-ion collisions. 
In the expanding case, more and more modes become unstable as
the expansion increases the momentum anisotropy, while the growth
rate of each unstable mode decreases with the density of the plasma. 
In \cite{Rebhan:2008uj} first numerical
results for the non-Abelian evolution were obtained, albeit
still restricted to the effectively 1+1-dimensional situation
of modes that are constant in transverse directions.
In this case, continued approximately exponential growth was observed (albeit
only exponential in the square root of proper time due to the
linearly decreasing density of the plasma).

In the expanding case, even the time evolution in the
weak-field regime is
nontrivial and cannot be given in closed form.
However, in Ref.~\cite{Romatschke:2006wg} integro-differential
equations were obtained which made it possible to study
the time evolution semi-analytically in the case of
effectively 1+1-dimensional dynamics.
As a preparation of fully 3+1-dimensional real-time lattice
studies of non-Abelian plasma instabilities in expanding plasmas,
we shall generalize the semi-analytical treatment of
Ref.~\cite{Romatschke:2006wg} to arbitrary wave vectors of
the collective modes. 

At the same time, we shall generalize
to arbitrary initial conditions, allowing for seeds
both in the collective gauge fields (the case considered
in \cite{Romatschke:2006wg,Rebhan:2008uj}) and also in
the auxiliary fields corresponding more directly to colored fluctuations
in the hard particle distribution. 
While in the case of stationary anisotropic plasmas little difference
was found between the two possibilities for introducing
seed fields for instabilities, in the expanding case we find that
this generalization reduces dramatically the uncomfortably
long delay of the onset of growth that was observed
in \cite{Romatschke:2006wg,Rebhan:2008uj}.
As will be discussed in more detail below, this in fact reverts
some of the negative conclusions of Ref.~\cite{Romatschke:2006wg,Rebhan:2008uj}
concerning the possible role of plasma instabilities in heavy-ion collisions
for explaining very early thermalization. While at RHIC energies
with parameters matched from color glass condensate scenarios
\cite{Iancu:2003xm}, there is still somewhat too little room for
plasma instabilities that grow from small initial rapidity fluctuations,
the situation at LHC energies, which appeared somewhat
marginal in Ref.~\cite{Romatschke:2006wg,Rebhan:2008uj},
is now much more favorable with regard to an important role
of such plasma instabilities if the quark-gluon plasma to be produced
at the LHC is described by the weak-coupling physics underlying
the hard-loop-resummed treatment.

\section{Hard-loop effective field equations with anisotropically expanding background}

A (sufficiently small)
gauge coupling $g$ introduces a 
hierarchy of smaller momentum scales below 
the scale of ``hard'' momenta $|\mathbf p|=p^0$ of ultrarelativistic 
plasma constituents.
The ``soft'' 
scale $\sim g\sqrt{f}\,|\mathbf p|$, where $f$ is the typical hard
particle occupation number, is
associated with different screening phenomena
and the various branches of plasmon propagation. To leading order
they are described by hard (thermal) loop effective theories.
When $f(\mathbf p)$ is anisotropic, the soft scale is also the domain
of plasma instabilities, which constitute
the dominant nonequilibrium effects at weak coupling: the associated rates are
parametrically
larger than any of the scattering processes among plasma
particles.

As long as the amplitude of the
gauge fields $A\ll \sqrt{f}\,|\mathbf p|$, the evolution of the plasma instabilities
is essentially Abelian and can be studied by a perturbative linear response analysis.
For a stationary anisotropic plasma, the evolution is simply
exponential in time
until non-Abelian self-interactions might hinder further
growth when $A\gtrsim \sqrt{f}\,|\mathbf p|$
and thereby delay the isotropization coming from
the backreaction of the collective fields on the
distribution of hard plasma particles.
In an expanding plasma, the Abelian (weak field) regime
is complicated by the counterplay of increasing anisotropy, which
favors the appearance of plasma instabilities, and dilution
of hard particle densities as well as energy densities in soft collective
fields.  

In the following we shall first recapitulate the hard-loop effective field
equations for an anisotropically expanding non-Abelian plasma
as introduced in Refs.\ \cite{Romatschke:2006wg,Rebhan:2008uj}, and
later specialize to the effectively Abelian weak-field regime. 
In view of the numerical simulations that were carried out
to study the non-Abelian, strong-field regime, the Abelian evolution
can be expected to provide an upper limit on the strength
of plasma instabilities that are seeded by small initial fluctuations. 

\subsection{Boost invariant expanding background}

We assume a color-neutral background distribution of plasma particles
$f_0(\1p,\1x,t)$
which satisfies
\be\label{vdf0}
v\cdot \6\, f_0(\1p,\1x,t)=0,\qquad v^\mu=p^\mu/p^0.
\ee
This is satisfied trivially in a stationary homogeneous plasma with
arbitrary momentum anisotropy. In order to describe the earliest
stage of heavy-ion collisions in the limit of large nuclei, we
consider a plasma that expands in one spatial direction (the beam axis).
Requiring boost invariance
\cite{Bjorken:1983qr}
and isotropy in the transverse directions, we are led to
\cite{Baym:1984np,Mueller:1999pi}
\be
f_0(\mathbf p,x)=f_0(p_\perp,p^z,z,t)=f_0(p_\perp,p'^z,\tau)
\ee
where the transformed longitudinal momentum is
\be
p'^z=\gamma(p^z-\beta p^0),\quad
\beta=z/t,\quad \gamma=t/\tau,\quad \tau=\sqrt{t^2-z^2},
\ee
with $p^0=\sqrt{p_\perp^2+(p^z)^2}$ for ultrarelativistic (massless) particles.

Introducing the comoving coordinates of proper time $\tau$
and spacetime rapidity $\eta$ through
\bea
t=\tau \cosh\eta,\quad &&\beta=\tanh\eta,\nn\\
z=\tau \sinh\eta,\quad  &&\gamma=\cosh\eta,
\eea
we are led to a coordinate system with nontrivial metric
\be
ds^2=g_{\alpha\beta}dx^\alpha dx^\beta\equiv
d\tau^2-d\mathbf x_\perp^2-\tau^2 d\eta^2,
\ee
where indices from the beginning of the Greek alphabet will
be reserved for the new coordinates. Latin indices will from now on only
refer to the two transverse spatial directions.
Despite the nontrivial metric, we shall only use
ordinary derivatives when writing for instance
$ D_\alpha=\6_\alpha-ig[ A_\alpha,\cdot]$ for
gauge-covariant derivatives. 

In addition to space-time rapidity $\eta$, we also introduce
momentum space rapidity $y$ for the massless particles according to
\be
p^\mu=p_\perp(\cosh y,\cos\phi,\sin\phi,\sinh y).
\ee
In comoving coordinates, we then have
\bea
p^\tau&=&\sqrt{p_\perp^2+\tau^2( p^\eta)^2}\nn\\
&=&p^0\cosh\eta -p^z\sinh\eta
=p_\perp \cosh(y-\eta),\\
p^\eta&=&- p_\eta/\tau^2=(p^z\cosh\eta-p^0\sinh\eta)/\tau\nn\\
&=&p'^z/\tau
=[p_\perp \sinh(y-\eta)]/\tau.\label{peta}
\eea

Eq.~(\ref{vdf0}), where the space-time derivatives act at fixed
$\mathbf p_\perp$ and $p^z$, becomes
\be
( p^\alpha\cdot  \6_\alpha) f_0\Big|_{p^\mu}=0
\ee
with fixed $p^\mu$ and thus fixed
$p_\perp,y,\phi$ as opposed to fixed $p^\alpha$.
This is solved by any function
of the form $f_0(\1p,\1x,t)=f_0(\1p_\perp, p_\eta)$, taking into account
that $p_\eta$ depends on $\tau$ and $\eta$ according to (\ref{peta}).
We choose
\be\label{faniso}
f_0(\mathbf p,x)=f_{\rm iso}\left(\sqrt{p_\perp^2+ p_\eta^2/\tau_{\rm iso}^2}\right)=f_{\rm iso}\left(\sqrt{p_\perp^2+({p'^z\tau\0\tau_{\rm iso}})^2}\right)
\ee
which corresponds to a locally isotropic distribution on the hypersurface $\tau=\tau_{\rm iso}$ with increasingly oblate momentum space anisotropy at $\tau>\tau_{\rm iso}$ but prolate anisotropy for $\tau<\tau_{\rm iso}$.\footnote{Notice
that $\tau_{\rm iso}$ is just a parameter of the background distribution and
does not refer to the time where isotropization of the plasma
eventually occurs through interactions.}
We shall also have to consider a lowest value of proper time, $\tau_0$,
where a particle description of the plasma constituents begins to make
sense. Depending on whether the parameter $\tau_{\rm iso}$ is smaller or larger
than $\tau_0$ we shall consider a plasma that starts with oblate or
prolate momentum distribution.

The particle distribution function (\ref{faniso}) has the same form as the one used in 
Refs.~\cite{Romatschke:2003ms,Romatschke:2004jh,Rebhan:2004ur,Rebhan:2005re},
but the anisotropy parameter $\xi$ introduced therein
is now space-time dependent
according to 
\be\label{xi}
\xi(\tau)=(\tau/\tau_{\rm iso})^2-1,
\ee
and the
normalization factor $N(\xi)$ of 
Ref.~\cite{Romatschke:2004jh,Rebhan:2004ur,Rebhan:2005re}
is unity. (The anisotropy parameter
$\theta$ used in Ref.~\cite{Arnold:2007cg} 
is related to $\xi$ by $\xi\sim\theta^{-2}$ for large anisotropies.)
The behavior $\xi\sim\tau^2$ at large $\tau$ corresponds to
having a free-streaming background distribution. In a more realistic
collisional plasma, $\xi$ will have to grow slower than this.
In the first stage of the original bottom-up scenario \cite{Mueller:2005un}, ignoring
plasma instabilities, one would have had $\xi\sim\tau^{2/3}$. 
In Ref.~\cite{Bodeker:2005nv} it was argued that plasma instabilities
reduce the exponent to $\xi\sim\tau^{1/2}$, whereas Ref.~\cite{Arnold:2007cg}
recently presented arguments in favor of $\xi\sim\tau^{1/4}$.
All these scenarios typically
consider $\xi\gg1$, so below we shall mostly concentrate
on the case $\tau_{\rm iso}<\tau_0$ and thus high anisotropy for all $\tau>\tau_0$,
but in the simplified case of a collisionless free-streaming expansion.
However, we shall also discuss instabilities in prolate phases
for which we need to set $\tau_{\rm iso}>\tau_0$.

\subsection{Hard-expanding-loop effective field equations}

In an approximately collisionless plasma,
the gauge covariant Boltzmann-Vlasov equations for color charge
carrying perturbations $\delta f_a$ have the
form
\be\label{vDf}
v\cdot D\, \delta f_a(\1p,\1x,t)=g v_\mu F^{\mu\nu}_a \6^{(p)}_\nu f_0(\1p,\1x,t).
\ee
In comoving coordinates we write
\be\label{VDf}
 V\cdot  D\, \delta f^a\big|_{p^\mu}=g  V^\alpha 
 F_{\alpha\beta}^a  \6_{(p)}^\beta f_0(\1p_\perp, p_\eta),
\ee
where in place of the light-like vector $v^\mu=p^\mu/p^0$ 
containing a unit 3-vector we introduced the
new quantity 
\be
 V^\alpha = { p^\alpha \0 p_\perp} =
\left(\cosh (y-\eta),\,\cos\phi,\,\sin\phi,\,{1\0\tau}\sinh (y-\eta)\right),
\label{velocityDef1}
\ee
which is normalized so that it has a unit 2-vector in the transverse plane.

Eq.~(\ref{VDf}) can be solved in terms of an auxiliary field
$W_\beta( x;\phi,y)$ which satisfies
\be\label{VDW}
 V\cdot  D\,  W_\beta\big|_{\phi,y}= V^\alpha  F_{\beta\alpha} \, ,
\ee
and
\be\label{Wdef}
\delta f(x;p)=-g  W_\beta( x;\phi,y) 
\6_{(p)}^\beta f_0(p_\perp, p_\eta).
\ee

Expressed in terms of the auxiliary field $ W$, 
the induced current in comoving
coordinates reads
\bea\label{tjind}
 j^\alpha &=& g t_R \int {d^3p\over(2\pi)^3 2p^0}p^\alpha
\delta f(x;p) \nn\\
&=&-g^2t_R
\int_0^\infty {p_\perp dp_\perp \over 8\pi^2}
\int_0^{2\pi} {d\phi \0 2\pi}
\int_{-\infty}^\infty dy \,
 p^\alpha\, {\partial f_0 
\over \partial  p_\beta}
 W_\beta (x;\phi,y) \, ,
\eea
where $t_R$ is a suitably normalized group factor.
Now for each $(\phi,y)$ (i.e., fixed $\1v$) the scale
$p_\perp$ (related to energy by $p^0=p_\perp\cosh y$) can be integrated out.

With the particular background distribution function (\ref{faniso})
we obtain
\be
 j^\alpha=-m_D^2\, \2\int_0^{2\pi} {d\phi \0 2\pi}
\int_{-\infty}^\infty dy \, 
 V^\alpha
\left(1+{\tau^2\0\tau_{\rm iso}^2}\sinh^2(y-\eta)\right)^{-2}
\left(V^{\9{i}} W_{\9{i}}+{\tau^2\0\tau_{\rm iso}^2} V^\eta\,  W_\eta\right)\, ,
\label{current}
\ee
where
\be
m^2_D=-g^2t_R \int_0^\infty {dp\,p^2\0(2\pi)^2} f'_{\rm iso}(p) 
\ee
equals the Debye mass squared at the 
time $\tau_{\rm iso}$ (which we shall often choose smaller than $\tau_0$
so that $m_D$ is not physically realized by just a convenient mass parameter). 

These equations are closed by the non-Abelian Maxwell equations which
in comoving coordinates read
\be
{1\0\tau} D_\alpha(\tau  F^{\alpha\beta}) \equiv
{1\0\tau} D_\alpha\left[\tau g^{\alpha\gamma}(\tau) g^{\beta\delta}(\tau)
 F_{\gamma\delta}\right]= j^\beta,
\ee
where $ F_{\alpha\beta}=\6_\alpha  A_\beta
-\6_\beta  A_\alpha
-ig[ A_\alpha, A_\beta]$.
To solve them, we adopt the comoving
temporal gauge $A^\tau=0$ and introduce canonical conjugate field momenta
for the remaining gauge fields according to\footnote{In the following
field equations we keep the index position of conjugate field momenta opposite
to that of the associated fields.}
\be\label{Piidef}
\Pi^{\9{i}} = \tau \6_\tau A_{\9{i}} = 
-\tau \6_\tau A^{\9{i}}=-\Pi_{\9{i}} \, ,
\ee
and
\be\label{Pietadef}
\Pi^\eta={1\0\tau}\6_\tau A_\eta \, .
\ee

In terms of fields and conjugate momenta,
the Yang-Mills field equations then read
\bea\label{YMeqs}
\tau \6_\tau \Pi^\eta&=&j_\eta-D_{\9{i}} F^{\9{i}}{}_\eta\,,\\
\tau^{-1}\6_\tau \Pi_{\9{i}}&=&j^{\9{i}}
-D_{\9{j}}F^{{\9{j}}{\9{i}}}
-D_\eta F^{\eta{\9{i}}}\,.
\eea
In a comoving frame, the longitudinal (chromo-)electric and magnetic fields
are given by 
\be
E^\eta=\Pi^\eta,\quad B_\eta=F_{12}, 
\ee
but
transverse components involve a factor of $\tau$,
\be
E^{\9{i}}=\tau^{-1} \Pi^{\9{i}},\quad
B_i=\tau^{-1} F_{\eta j}\epsilon_{ji}.
\ee
In terms of these, the contribution to the energy density
is simply
\be
\mathcal E = \mathcal E_{T}+\mathcal E_{L}= \mathcal E_{B_T}+\mathcal E_{E_T} +\mathcal E_{B_L}+\mathcal E_{E_L} =
\tr \left[ (B_i)^2 + (E^i)^2 + (B_\eta)^2 + (E^\eta)^2 \right].
\ee
However, due to the expansion, the total energy density $\mathcal E$ is not conserved, even when the
induced current (\ref{tjind}) is identically zero,
\be
\frac{d}{d\tau}\mathcal E|_{j\equiv0}=-\frac{2}{\tau}\mathcal E_T|_{j\equiv0}\,.
\ee

In the presence of a plasma of hard particles and thus nonvanishing
induced current $j$ we define the net energy gain rate by
\be\label{REG}
R_{\,\rm Gain} \; \equiv \; 
  \frac{d{\cal E}}{d\tau} + \frac{2}{\tau}{\cal E}_T\,,
\ee
which gives the rate of energy transfer
from the free-streaming hard 
particles into the collective chromo-fields and which is positive when
plasma instabilities are at work. For stable modes $R_{\,\rm Gain}$ 
oscillates about zero.

\section{Time evolution of gauge fields in the weak-field
regime}

In the regime where self interactions of gauge fields cannot be
neglected, the Yang-Mills field equations and the equations of motion
for the $W_\alpha(x;\phi,y)$ fields are nonlinear and require
numerical, real-time lattice evaluation. In the limit of
small field amplitudes, these equations are linear and can be
reduced to ordinary integro-differential equations in proper time
for individual modes.

\subsection{Solving the $W$ field equations}

In the weak-field regime, the $W$ field equations reduce to
\be
(V^\tau \6_\tau+V^i \6_i+V^\eta \6_\eta)
W_\alpha(\tau,x^i,\eta;\phi,y)=V^\beta F_{\alpha\beta}
=V^\beta(\6_\alpha A_\beta-\6_\beta A_\alpha).
\ee
Because our background distribution function is isotropic in
the transverse plane, we can without loss of generality restrict to
modes which are independent of $x^{i=2}$, keeping only $x^{i=1}\equiv x$
(recall that the symbol $y$ is already used for momentum rapidity).
In temporal axial gauge we thus have
\bea\label{W2+1}
(V^\tau \6_\tau+V^\eta \6_\eta+\cos\phi\, \6_x)
W_\alpha(\tau,x,\eta;\phi,y)
&=&-V^\tau \6_\tau A_\alpha+V^\eta(\6_\alpha A_\eta-\6_\eta A_\alpha)\nn\\
&&+\cos\phi\;(\6_\alpha A_1 - \6_x A_\alpha)+\sin\phi\; \6_\alpha A_2\,.
\eea
These first-order partial differential equations can be solved by the
method of characteristics as follows. We introduce a parameter $s$
such that the left-hand side of Eq.~(\ref{W2+1}) is replaced by
\be
\frac{d W_{\alpha}}{d s}=\frac{\partial W_{\alpha}}{\partial \tau} \frac{d \tau}{d s}+\frac{\partial W_{\alpha}}{\partial \eta} \frac{d \eta}{d s}
+\frac{\partial W_{\alpha}}{\partial x} \frac{d x}{d s}
\ee
with
\bea
\frac{d \tau}{d s}&=&V^{\tau}=\cosh(y-\eta(s)),\\
\frac{d \eta}{d s}&=&V^{\eta}=\frac1{\tau(s)}\sinh(y-\eta(s)),\\
\frac{d x}{d s}&=&V^1=\cos\phi\,.
\eea
Since ${d\tau}/{ds} > 0$ we can use $\tau$ in place of $s$ 
for the purpose of integrating Eq.~(\ref{W2+1}).
Writing $ds={d\tau'}/{V^{\tau}(\eta(\tau'))}$ we obtain
\begin{equation}
W_{\alpha}(\tau,x,\eta;\phi,y)-W_{\alpha}(\tau_{0},x_0,\eta_{0};\phi,y)=
\int_{\tau_{0}}^{\tau} d \tau' \frac{V^{\beta}F_{\alpha \beta}|_{\tau',x(\tau'),\eta(\tau')}}{\cosh(y-\eta(\tau'))}
\label{eq:W}
\end{equation}
with $x_0\equiv x(\tau'=\tau_0)$ and $\eta_{0}\equiv \eta(\tau'=\tau_{0})$.
The functions $\eta(\tau')$ and $x(\tau')$ are solutions of
\bea
\label{etataueq}
\frac{d \eta(\tau')}{d \tau'}&=&\frac{1}{\tau'}\tanh(y-\eta(\tau'))\\
\label{xtaueq}
\frac{d x(\tau')}{d \tau'}&=&\frac{\cos\phi}{\cosh(y-\eta(\tau'))}
\eea
with initial conditions $\eta(\tau'=\tau)=\eta$ and $x(\tau'=\tau)=x$.
Eq.\ (\ref{etataueq}) is solved by \cite{Romatschke:2006wg}
\be
\tau'\sinh(y-\eta(\tau'))=\tau\sinh(y-\eta)
\ee
or, more explicitly,
\be\label{etapr}
\eta'\equiv \eta(\tau')=y-\asinh\left(\frac{\tau}{\tau'}\sinh(y-\eta)\right).
\ee
With this solution, Eq.~(\ref{xtaueq}) can be integrated, yielding
\be\label{xpr}
x'\equiv x(\tau')=x+
\left[\tau'\cosh(y-\eta')-\tau\cosh(y-\eta)\right]\cos\phi.
\ee

The $W$ fields, from which the induced current is obtained upon integration
over $\phi$ and $y$ according to Eq.~(\ref{current}), are now given
explicitly by the following ``memory integrals''
\bea\label{memints}
W_1-W_1^0&=&\int_{\tau_0}^\tau d\tau'\left[\partial_{\tau'}A^1-\frac{\tanh(\eta'-y)}{\tau'}(\6_{x'} A_\eta+\6_{\eta'}A^1)-
{\sin\phi\0\cosh(\eta'-y)}\6_{x'} A^2\right],\nn\\
W_2-W_2^0&=&\int_{\tau_0}^\tau d\tau'\left[\partial_{\tau'}-\frac{\tanh(\eta'-y)}{\tau'}\6_{\eta'}+
{\cos\phi\0\cosh(\eta'-y)}\6_{x'}\right] A^2,\nn\\
W_\eta-W_\eta^0&=&-\int_{\tau_0}^\tau d\tau'\left[\partial_{\tau'}A_\eta
+\frac{V^i\6_{\eta'}A^i+\6_{x'}A_\eta}{\cosh(\eta'-y)}\right],
\eea
where inside the integrals $A_\alpha=A_\alpha(\tau',x',\eta')$.
Note that $x'$ and $\eta'$ as well as 
$x_0$ and $\eta_0$ appearing as arguments
in $W_{\alpha}^0\equiv W_{\alpha}(\tau_{0},x_0,\eta_{0};\phi,y)$
all are functions of the spacetime variables $\tau$, $\eta$ and $x$. 

\subsection{Fourier components}\label{sec:Fc}

Because of the linearity of the Maxwell (linearized Yang-Mills) equations and the $W$ equations in the weak-field (Abelian) regime and their translational invariance
in $\eta$ and $y$, we can study the time evolution of individual
modes obtained by a Fourier decomposition
\be
A_\alpha(\tau,x,\eta)=\int\frac{dk}{2\pi}e^{i kx}\!\!\int\frac{d\nu}{2\pi}e^{i \nu \eta}
\tilde{A}_\alpha(\tau;k,\nu).
\ee
With similarly Fourier transformed currents, the equations of motion 
for $\tilde A^1$ and $\tilde A_\eta$ read
\bea\label{tiAeta1}
(\tau^{-1}\6_\tau \tau \6_\tau + \tau^{-2}\nu^2)\tilde A^1(\tau;k,\nu)
&=&\tilde j^1(\tau;k,\nu)-k\nu\tau^{-2}\tilde A_\eta(\tau;k,\nu)\nn\\
(\tau\6_\tau \tau^{-1} \6_\tau + k^2)\tilde A_\eta(\tau;k,\nu)
&=&\tilde j_\eta(\tau;k,\nu)-k\nu \tilde A^1(\tau;k,\nu)
\eea
with $\tilde j^1(\tau;k,\nu)$ and $\tilde j_\eta(\tau;k,\nu)$ both depending
on $\tilde A^1(\tau';k,\nu)$ and $\tilde A_\eta(\tau';k,\nu)$, $\tau'\le
\tau$, but not on $\tilde A^2(\tau';k,\nu)$, as we shall see shortly.
The equation of motion for the transverse mode
$\tilde A^2(\tau;k,\nu)$ is decoupled
from $\tilde A^1$ and $\tilde A_\eta$ and reads
\be\label{tiA2}
(\tau^{-1}\6_\tau \tau \6_\tau + k^2+\tau^{-2}\nu^2)\tilde A^2(\tau;k,\nu)
=\tilde j^2(\tau;k,\nu)
\ee
with $\tilde j^2(\tau;k,\nu)$ depending on the history of $\tilde A^2(\tau';k,\nu)$.

In order to express the current as functional of the gauge fields,
we first note that in Eqs.~(\ref{memints}),
the partial derivatives $\6_{x'}$ and $\6_{\eta'}$ are simply replaced
by factors $ik$ and $i\nu$, respectively, while the proper-time integrals
over the {\em partial} time derivative of the gauge
fields $A_\alpha(\tau',x',\eta')$ can be integrated by parts, yielding
\bea\label{dtaupAint}
\int_{\tau_0}^\tau d\tau'\partial_{\tau'}A_\alpha(\tau',x',\eta')&=&
\int_{\tau_0}^\tau d\tau'
\int\frac{dk}{2\pi}e^{i kx'(\tau')}\!\!\int\frac{d\nu}{2\pi}e^{i \nu \eta'(\tau')}
\partial_{\tau'}\tilde{A}_\alpha(\tau;k,\nu)\\
&=&\int\frac{dk}{2\pi}\!\!\int\frac{d\nu}{2\pi}\biggl\{
e^{ikx}e^{i\nu\eta}\tilde{A}_\alpha(\tau;k,\nu)
-e^{ikx_0(\tau)}e^{i\nu\eta_0(\tau)}\tilde{A}_\alpha(\tau_0;k,\nu)\nn\\
&&-i\int_{\tau_0}^\tau d\tau'
e^{i kx'}e^{i \nu \eta'}\left[
k {\cos\phi\0\cosh(y-\eta')}+\nu\frac{\tanh(y-\eta')}{\tau'}\right]
\tilde{A}_\alpha(\tau';k,\nu)\biggr\}.\nn
\eea

Inserting Eqs.~(\ref{memints}) with 
(\ref{dtaupAint}) into the expression for the current,
Eq.~(\ref{current}) and introducing the abbreviations\footnote{The
notation is chosen such that a bar indicates a dependence on $\bar y$ and
a prime a dependence on $\tau'$ ($\bar y$ and $\tau'$ are the
two remaining integration variables in Eqs.~(\ref{tij1})--(\ref{tijtau})).}
\bea
\bar y &\equiv& y-\eta, \nn\\
\bar\eta'&\equiv&\bar\eta(\tau')\equiv\eta(\tau')-\eta=\bar{y}-\asinh(\frac{\tau}{\tau'} \sinh\bar{y}),
\quad \bar\eta_0\equiv \bar\eta(\tau_0),\nn\\
\bar\chi'&\equiv&\bar\chi(\tau')\equiv(x(\tau')-x)/\cos\phi=\sqrt{\tau'^2+\tau^2\sinh^2\bar{y}}-\tau \cosh\bar{y},\quad \bar\chi_0\equiv
\bar\chi(\tau_0),
\eea
we find the following results for the Fourier components
$\tilde j^\alpha(\tau;k,\nu)$
after performing the integration over
the (momentum space) angle $\phi$ 
in terms of Bessel functions $J_n$,
 \begin{eqnarray}\label{tij1}
  \tilde{j}^1&=&-\frac{m_D^2}{2}\!\int\!\frac{d\bar{y}}{\bigl(1+\frac{\tau^2 \sinh^2\bar{y}}{\tau_{\rm iso}^2}\bigr)^2}\Bigg\{\frac{1}{2}\tilde{A}^1(\tau)+e^{i \nu\bar{\eta}_0}\Bigg[\frac{i \tau\sinh\bar{y}}{\tau_{\rm iso}^2}J_1(k\bar\chi_0)\tilde{A}_\eta(\tau_0)
-\frac{1}{2}[J_0-J_2](k\bar\chi_0)\tilde{A}^1(\tau_0)\Bigg]\nonumber\\
  &&\qquad+\int^{\tau}_{\tau_0}d\tau' \frac{e^{i\nu \bar{\eta}'}}{\sqrt{1+\frac{\tau^2 \sinh^2\bar{y}}{\tau'^2}}}\Bigg[\Big(\frac{k}{4}[3J_1-J_3](k\bar\chi')-\frac{i\nu\tau\sinh\bar{y}}{2\tau_{\rm iso}^2}[J_0-J_2](k\bar\chi')\Big)\tilde{A}^1(\tau')\nonumber\\
  &&\qquad\qquad+\frac{\tau\sinh\bar{y}}{\tau'^2}\Big(\frac{ik}{2}[J_0-J_2](k\bar\chi')-\frac{\nu \tau\sinh\bar{y}}{\tau_{\rm iso}^2}J_1(k\bar\chi')\Big)\tilde{A}_\eta(\tau')\Bigg]\Bigg\}+\tilde j^1_0(\tau),
 \end{eqnarray} 
and
 \begin{eqnarray}\label{tijeta}
  \tilde{j}^\eta&=&-\frac{m_D^2}{2\tau}\!\int\!\frac{d\bar{y}\sinh\bar{y}}{\bigl(1+\frac{\tau^2 \sinh^2\bar{y}}{\tau_{\rm iso}^2}\bigr)^2}\Bigg\{-\frac{\tau\sinh\bar{y}}{\tau_{\rm iso}^2}\tilde{A}_\eta(\tau)
+e^{i\nu\bar{\eta}_0}\Bigg[\frac{\tau\sinh\bar{y}}{\tau_{\rm iso}^2}J_0(k\bar\chi_0)\tilde{A}_\eta(\tau_0)-i J_1(k\bar\chi_0)\tilde{A}^1(\tau_0)\Bigg]\nonumber\\
  &&\qquad+\int^{\tau}_{\tau_0}d\tau' \frac{e^{i\nu \bar{\eta}'}}{\sqrt{1+\frac{\tau^2 \sinh^2\bar{y}}{\tau'^2}}}\Bigg[\Big(\frac{ik}{2}[J_2-J_0](k\bar\chi')+\frac{\nu \tau\sinh\bar{y}}{\tau_{\rm iso}^2}J_1(k\bar\chi')\Big)\tilde{A}^1(\tau')\nonumber\\
  &&\qquad\qquad+\frac{\tau\sinh\bar{y}}{\tau'^2}\Big(\frac{i \nu \tau\sinh\bar{y}}{\tau_{\rm iso}^2}J_0(k\bar\chi')-kJ_1(k\bar\chi')\Big)\tilde{A}_\eta(\tau')\Bigg]\Bigg\}+\tilde j^\eta_0(\tau).
 \end{eqnarray}
As mentioned above, these components of the current only depend on (the
history of) $\tilde A^1$ and $\tilde A_\eta$. 
If either $k=0$ or $\nu=0$, the 1 and $\eta$ components decouple
from each other. (For $k=0$ only the terms involving $J_0(0)=1$ survive,
while for $\nu=0$ all terms involving odd powers of $\sinh\bar y$
integrate to zero.)
On the other hand,
$\tilde{j}^2$ is found to be a functional of only $\tilde A^2$,
\bea\label{tij2}
  \tilde{j}^2&=&-\frac{m_D^2}{4}\!\int\!\frac{d\bar{y}}{\bigl(1+\frac{\tau^2 \sinh^2\bar{y}}{\tau_{\rm iso}^2}\bigr)^2}\Bigg\{\tilde{A}^2(\tau)-e^{i \nu\bar{\eta}_0}[J_0+J_2](k\bar\chi_0)\tilde{A}^2(\tau_0)\\
  &&\qquad+\int^{\tau}_{\tau_0}d\tau' \frac{e^{i\nu \bar{\eta}'}}{\sqrt{1+\frac{\tau^2 \sinh^2\bar{y}}{\tau'^2}}}\Big(\frac{k}{2}[J_1+J_3](k\bar\chi')-\frac{i\nu\tau\sinh\bar{y}}{\tau_{\rm iso}^2}[J_0+J_2](k\bar\chi')\Big)\tilde A^2(\tau')\Bigg\}+\tilde j^2_0(\tau).\nn
\eea 

Eqs.~(\ref{tij1})--(\ref{tij2}) generalize the expressions given in \cite{Romatschke:2006wg}
for the effectively 1+1-dimensional
case $k=0$, where the wave vector of the collective modes
points in the direction of momentum-space anisotropy. (As they should, 
the functional dependences in $\tilde j^1[\tilde A^1]$
and $\tilde j^2[\tilde A^2]$ become the same for $k=0$.)

The $\tau$ component of the current, which is needed only for
a check of the Gauss law constraint, is again independent of
$\tilde A^2$ and given by the following functional of
$\tilde A^1$ and $\tilde A_\eta$,
 \begin{eqnarray}\label{tijtau}
 \tilde{j}^\tau&=&-\frac{m_D^2}{2}\!\int\!\frac{d\bar{y} \cosh\bar{y}}{\bigl(1+\frac{\tau^2 \sinh^2\bar{y}}{\tau_{\rm iso}^2}\bigr)^2}\Bigg\{e^{i \nu \bar{\eta}_0}\Bigg[\frac{\tau\sinh\bar{y}}{\tau_{\rm iso}^2}J_0(k\bar\chi_0)\tilde{A}_\eta(\tau_0)-i J_1(k\bar\chi_0)\tilde{A}^1(\tau_0)\Bigg] \nn\\
 &&\qquad+\int^{\tau}_{\tau_0}d\tau' \frac{e^{i\nu \bar{\eta}'}}{\sqrt{1+\frac{\tau^2 \sinh^2\bar{y}}{\tau'^2}}}\Bigg[\Big(\frac{i k}{2}[J_2-J_0](k\bar\chi')+\frac{\nu \tau\sinh\bar{y}}{\tau_{\rm iso}^2}J_1(k\bar\chi')\Big)\tilde{A}^1(\tau') \nonumber\\
 &&\qquad\qquad+\frac{\tau\sinh\bar{y}}{\tau'^2}\Big(\frac{i \nu \tau\sinh\bar{y}}{\tau_{\rm iso}^2}J_0(k\bar\chi')-kJ_1(k\bar\chi')\Big)\tilde{A}_\eta(\tau')\Bigg]\Bigg\}+\tilde j^\tau_0(\tau).
 \end{eqnarray} 

In all these expressions, $\tilde j^\alpha_0(\tau;k,\nu)$ corresponds
to nontrivial initial data for $W_\alpha$, if any. Up to a factor
$e^{ikx}e^{i\nu\eta}$, the Fourier component
$\tilde j^\alpha_0(\tau;k,\nu)$ is
obtained by evaluating the expression for the induced 
current, Eq.~(\ref{current}),
with $W$'s of the form
\be\label{tiW0}
W_{\alpha}(\tau_{0},x_0(\tau),\eta_{0}(\tau);\phi,y)
=e^{ikx_0(\tau)}e^{i\nu \eta_0(\tau)}
\tilde W^0_\alpha(k,\nu;\phi,y-\eta_0(\tau)),
\ee
where the functions $x_0$ and $\eta_0$ are given by
Eqs.~(\ref{xpr}) and (\ref{etapr}), respectively, with
$\tau'=\tau_0$.
Nontrivial initial values $\tilde W^0_\alpha$ are
required whenever the right-hand side of the Gauss law constraint
at $\tau=\tau_0$,
\be\label{GLconstr}
\tilde j^\tau(\tau_0)=\frac{i}{\tau_0}\left(\nu\tilde\Pi^\eta(\tau_0)
+k\tilde \Pi^1(\tau_0) \right),
\ee
is nonvanishing. This is naturally the case for modes where the polarization
of the gauge fields and their momenta (in temporal gauge)
is longitudinal with respect to the spatial wave vector.

Note that already for a single mode
there is considerable freedom in the choice of
initial conditions
which is parametrized by the functions 
$\tilde W^0_\alpha(k,\nu;\phi,y-\eta_0)$
and which will be explored in detail below.

\subsection{Stable and unstable modes}\label{sect:modes}

Before studying the time evolution of the individual Fourier modes
through numerical evaluation of the integro-differential equations
provided by the above expressions, it
is useful to recall  
the case of a stationary anisotropic plasma, where a rather complete analysis of stable and unstable modes in the regime of weak fields has been carried out in
Refs.~\cite{Romatschke:2003ms,Arnold:2003rq,Romatschke:2004jh}. 
This is still relevant for the expanding case, which is however complicated by the time dependence of the density of the plasma and its anisotropy parameter so that some unstable modes may shut off as the plasma evolves towards a higher degree of oblateness, while new ones come into being. 
Since the growth rate of unstable modes also depends on the orientation of the wave vector, a further complication comes from the fact that this orientation
is time dependent unless the wave vector is strictly parallel or orthogonal to the anisotropy direction. When both $k\not=0$ and $\nu\not=0$,
in comoving Cartesian coordinates the wave vector rotates into
the transverse plane according to
\be
\mathbf k=k \,\mathbf e_1 + \frac{\nu}{\tau}\,\mathbf e_3
\ee
with the angle between $\mathbf k$ and $\mathbf e_3$ given by
\be
\vartheta = \arctan\frac{\tau k}{\nu}.
\ee

When the anisotropy is characterized by only one spatial 
direction,
as is the case for the distribution (\ref{faniso}), there are
in general three different branches\footnote{For a given wave vector, one branch may have more than one mode, e.g., a propagating wave and a growing unstable mode.} of modes.
Following
Ref.~\cite{Romatschke:2003ms}, the modes with a polarization
of the electric field
transverse to
both wave vector and the direction of anisotropy 
are denoted by the label $\alpha$. In our case these
correspond to the Fourier components $\tilde A^2(\tau;k,\nu)$
and $\tilde j^2(\tau;k,\nu)$, described by Eqs.~(\ref{tiA2})
and (\ref{tij2}). 
Such modes are stable in the case of a prolate momentum anisotropy,
and unstable for all orientations of the wave vector in the case
of oblate anisotropy, but the growth rate of the latter
approaches zero as $\vartheta\to\frac{\pi}2$. In the terminology of
Ref.~\cite{Arnold:2003rq}, these instabilities are magnetic ones
and have been first described by Weibel \cite{Weibel:1959}.
The instabilities studied so far for expanding plasmas \cite{Romatschke:2006wg,Rebhan:2008uj}
correspond to such modes in the special case $\vartheta=0$.

\begin{table*}
\caption{Classification of modes according to Ref.~\cite{Romatschke:2003ms},
the corresponding polarization of the electric and magnetic field, and
the range of instabilities in terms of the angle $\vartheta$ between
the wave vector $\mathbf k$ and the direction of anisotropy.
\label{table1}}
\begin{ruledtabular}
\begin{tabular}{lllll}
Mode & E field  & B field  & 
instabilities, prolate case & instabilities, oblate case \\ \hline
$\alpha$ & $\mathbf k \perp \mathbf E \parallel \mathbf e_2$ & $\mathbf k \perp \mathbf B \perp \mathbf e_2$ 
& stable & $0\le \vartheta < \frac{\pi}2$ \\
$+$ 
& $\mathbf E \perp \mathbf e_2$ & $\mathbf k \perp \mathbf B \parallel \mathbf e_2 $ 
& stable & stable \\ 
$-$ & $\mathbf E \perp \mathbf e_2$ & $\mathbf k \perp \mathbf B \parallel \mathbf e_2 $ 
& $\frac{\pi}4 \lesssim \vartheta \le \frac{\pi}2$ & $0\le \vartheta \lesssim \frac{\pi}4$ \\
\end{tabular}
\end{ruledtabular}
\end{table*}

When the polarization of
the electric field (and of the gauge field in temporal gauge)
lies in the plane spanned by the wave vector and the direction of
anisotropy, there are two modes, labelled ``+'' and ``$-$''
in Ref.~\cite{Romatschke:2003ms}, according to which one has the larger ($+$)
or smaller ($-$) zero-frequency 
mass squared in the stationary anisotropic case.
In the isotropic case, the ``$+$''~mode corresponds to electric Debye
screening at zero frequency and longitudinal plasmons above
the plasma frequency, whereas the ``$-$''~mode coincides with the
$\alpha$ mode (which in this case has zero screening mass to
leading order). In the anisotropic case, this degeneracy is lifted as soon
as $\vartheta\not=0$. Now the ``$-$''~mode is unstable 
for both oblate and prolate anisotropies, but in each case for
only a limited range of angles $\vartheta$, see Table~\ref{table1}.

For generic orientation of the wave vector, the ``$-$''~mode
involves also longitudinal electric fields and thus can correspond to
an electric instability according to the classification of
Ref.~\cite{Arnold:2003rq}. However,
when either $k=0$ ($\vartheta=0$) or $\nu=0$ ($\vartheta=\frac{\pi}2$),
the equations for $\tilde A^1$ and $\tilde A_\eta$, (\ref{tiAeta1}), (\ref{tij1}) and (\ref{tijeta}), decouple. The mode which is longitudinal with respect
to $\mathbf k$ is then electrically screened, whereas the other one is
transverse and corresponds
to either magnetostatic screening (when $\vartheta=\frac{\pi}2$
in the oblate case, or $\vartheta=0$ in the prolate case), or a
magnetic instability (when $\vartheta=0$
in the oblate case, or $\vartheta=\frac{\pi}2$ in the prolate case)\footnote{For large anisotropies, the magnetostatic screening mass may even become larger than the electrostatic (Debye) mass,\ e.g.\ with prolate anisotropy 
$\xi\lessapprox-0.88$, 
as is the case in Fig.~2b of Ref.~\cite{Romatschke:2003ms}, whereupon
electrostatic Debye screening at $\vartheta=0$ becomes part of the
``$-$'' branch and thus continuously connected with the magnetic
instability at $\vartheta=\frac{\pi}2$. 
}.

In the following we shall study the time evolution of
some representative cases of stable and unstable modes in the expanding case,
for initially oblate as well as prolate anisotropies,
with a particular view on
the dependence on initial conditions.

\section{Numerical evaluation}

The integro-differential equations for the time evolution of individual Fourier
modes that we have obtained in Sect.~\ref{sec:Fc} for the linear response regime
can be solved
straightforwardly by discretizing the proper time variables $\tau$
and employing a leap frog algorithm for gauge fields and conjugate momenta,
Eqs.~(\ref{Piidef})--(\ref{YMeqs}). The memory integrals in the expressions for
the induced currents also involve integrations over the momentum rapidity
variable $\bar y$ that have to be performed for each time step between
$\tau_0$ and $\tau$.

Since we are interested in the earliest stage of the evolution of
plasma instabilities from small initial fluctuations, we choose
our dimensionful quantities such that $\tau_0 \sim Q_s$, with $Q_s$
the so-called saturation scale of the color-glass-condensate (CGC) framework \cite{McLerran:1993ni,Iancu:2003xm},
and a normalization of the
hard particle distribution function (\ref{faniso}) such that
at $\tau=\tau_0$ the hard-gluon density of CGC estimates is matched.
This involves the so-called gluon liberation factor $c$
\cite{Baier:2002bt}, which
we choose as $c=2 \ln 2\approx 1.386$ as obtained in
approximate analytical calculations by Kovchegov 
in Ref.~\cite{Kovchegov:2000hz}. While this value is significantly
larger than the first numerical estimates \cite{Krasnitz:2001ph,Krasnitz:2003jw}, it is fairly
close to the most recent numerical result $c\simeq 1.1$
by Lappi \cite{Lappi:2007ku}. Our choice of $c$ corresponds to a value of the
squared mass parameter $m_D^2$ in the expression for the induced current
of $m_D^2\approx 1.285 (\tau_0 \tau_{\rm iso})^{-1}$.
(See Appendix C of Ref.~\cite{Rebhan:2008uj} for details.)
Unless stated otherwise, this value will be used in the following
numerical calculations. The only remaining free parameter is then
$\tau_{\rm iso}$, parametrizing the amount of anisotropy at the
initial time $\tau_0$, and we shall consider both initially
prolate and oblate distributions.

For later reference we note that
if we assume $\tau_0^{-1}\sim Q_s\sim 1$ and 3 GeV for RHIC and LHC experiments,
respectively, 1 fm/c corresponds to $\sim 5\tau_0$ for RHIC and $\sim 15 \tau_0$
for LHC. 

\subsection{Wave vector parallel to anisotropy direction}

Fourier modes with $k=0$ and $\nu\not=0$ have a wave vector parallel
to the spatial direction of momentum anisotropy and thus are
constant in the plane transverse to the axis of expansion. Such modes 
are stable for prolate anisotropy, while with oblate anisotropy
there are magnetic (Weibel) instabilities 
below a certain ($\xi$- or $\tau$-dependent) 
value of $\nu$. The case $k=0$, which has been studied
semi-analytically before in Ref.~\cite{Romatschke:2006wg},
is particularly interesting since it covers the most unstable mode of
a plasma with oblate anisotropy. 

Before studying the unstable modes in more detail
(eventually also with $k\not=0$), we begin with the
stable longitudinal modes, which are complicated by the need
for nonvanishing initial induced currents. The analysis of Ref.~\cite{Romatschke:2006wg} of
the unstable modes at $k=0$ will
then be generalized by allowing also for nonvanishing initial
currents.

\subsubsection{Stable (longitudinal plasmon) modes}

Longitudinal modes with $k=0$ and nonvanishing $A_\eta$, $\Pi^\eta$
are purely electrical and correspond to charge density waves (longitudinal
plasmons). 
Initial conditions only in $A_\eta$, with zero initial $\Pi^\eta$ and
$W$ fields would only yield a trivial, constant solution. 
For nontrivial solutions, we need nonvanishing
initial $W$ fields and nonzero initial currents. In order to
have nonzero $\tilde j^\tau(\tau_0;\nu)$ we need initial values
of the $W$ fields that are odd in $\bar y\equiv y-\eta$. A possible choice is
\begin{equation}
 \tilde W_\eta^0(k=0,\nu;\phi,y-\eta_0)=C_1 \tanh(y-\eta_0),
\label{eq:W_eta1}
\end{equation}
where $C_1$ is a constant. We recall that $\eta_0$ is given by (\ref{etapr}) as
\be
\eta_0\equiv
\eta(\tau'\!=\!\tau_0)=y-\asinh\left(\frac{\tau}{\tau_0}\sinh(y-\eta)\right)
\ee
so that $\eta_0=\eta$ at $\tau=\tau_0$.
The Gauss law constraint (\ref{GLconstr}) relates the constant $C_1$ to
the initial value of the longitudinal electric field according to
\begin{equation}
 \tilde{j}^\tau(\tau_0;\nu)=-\frac{2\pi C_1 m_D^2 \tau_0}{2\tau_{\rm iso}^2}\int \frac{d\bar{y} \sinh^2\bar{y}}{\left(1+\frac{\tau_0^2\sinh^2\bar{y}}{\tau_{\rm iso}^2}\right)^2}=\frac{i\nu\tilde{\Pi}^\eta(\tau_0;\nu)}{\tau_0}.
\end{equation}
The nonzero $W^0_\eta$ field (\ref{eq:W_eta1}) thus gives rise to
the following contributions $\tilde j_0^\alpha(\tau)$ in
the integral equations (\ref{tijeta}) and (\ref{tijtau}),
\bea
 \tilde{j}^\tau_0(\tau;\nu)&=&\frac{i\nu \tau^2\tilde{\Pi}^\eta(\tau_0;\nu)}{\tau_0^3}\,N^{-1}\int\frac{d\bar{y}\,e^{i\nu\bar{\eta}_0}\cosh\bar{y}\sinh^2\bar{y}}{\left(1+\frac{\tau^2\sinh^2\bar{y}}{\tau_{\rm iso}^2}\right)^2\sqrt{1+\frac{\tau^2\sinh^2\bar{y}}{\tau_0^2}}},\\
 \tilde{j}^\eta_0(\tau;\nu)&=&\frac{i\nu\tau \tilde{\Pi}^\eta(\tau_0;\nu)}{\tau_0^3}\,N^{-1}\int\frac{d\bar{y}\, e^{i\nu \bar{\eta}_0}\sinh^3\bar{y} }{\left(1+\frac{\tau^2\sinh^2\bar{y}}{\tau_{\rm iso}^2}\right)^2\sqrt{1+\frac{\tau^2\sinh^2\bar{y}}{\tau_0^2}}},
\eea
where\footnote{See Eq.~(\ref{Nanal}) for an analytic expression for $N$.}
\be
N(\tau_{\rm iso}/\tau_0)=\int\frac{d\bar{y} \sinh^2\bar{y}}{\left(1+\frac{\tau_0^2\sinh^2\bar{y}}{\tau_{\rm iso}^2}\right)^2}.
\ee
Recall that $\bar\eta_0\equiv \eta_0-\eta$, which vanishes at $\tau=\tau_0$,
so that $\tilde{j}^\eta_0(\tau_0;\nu)=0$.

\begin{figure}
 \centering
 \includegraphics[scale=1]{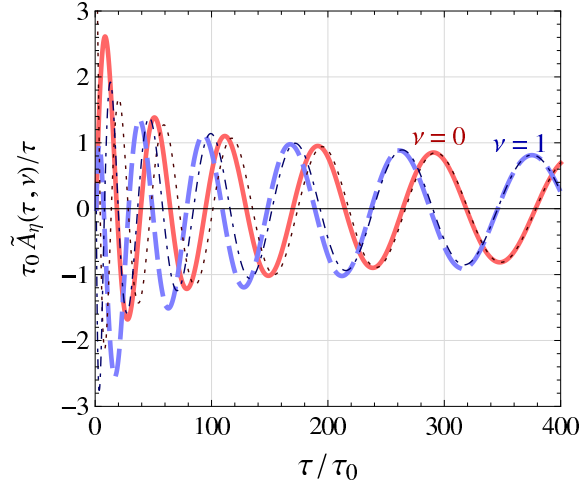}
 \caption{Longitudinal modes with $\nu=0$ and 1 for $\tau_{\rm iso}/\tau_0=10$. The dotted and dash-dotted lines represent the analytic results for the late time behavior, Eq.~(\ref{Aetaas}), which are in good agreement with the numerical data (thick full lines) at sufficiently large times.}
 \label{fig:long1}
\end{figure}

With $k=0$, the complete expression for the induced current (\ref{tijeta})
reads
\bea\label{tijetak0}
\tilde{j}^\eta(\tau;\nu)&=&\frac{m_D^2}{2\tau_{\rm iso}^2}\int\frac{d\bar{y}\sinh^2\bar{y}}{(1+\frac{\tau^2\sinh^2\bar{y}}{\tau_{\rm iso}^2})^2}\Biggl\{\tilde{A}_\eta(\tau;\nu)-e^{i\nu\bar{\eta}_0}\tilde{A}_\eta(\tau_0;\nu)\nonumber\\
&&\qquad-\int_{\tau_{0}}^\tau d\tau'\frac{i\nu\tau e^{i\nu\bar{\eta}'}\sinh\bar{y}}{\tau'^2\sqrt{1+\frac{\tau^2\sinh^2\bar{y}}{\tau_{\rm iso}}}}\tilde{A}_\eta(\tau';\nu)\Biggr\}+\tilde{j}^\tau_0(\tau;\nu),
\eea
which has to be solved together with
\begin{equation}
 \partial_\tau{1\0\tau}\6_\tau A_\eta(\tau;\nu)=-\tau\tilde{j}^\eta(\tau;\nu).
\end{equation}

In Appendix \ref{AppL} the late time behavior of the solutions is
derived in terms  of Bessel functions
$J_2$ and $Y_2$, see Eq.~(\ref{Aetaas}). 
In Fig.~\ref{fig:long1} this is compared with results of a full
numerical solution of the above integro-differential equation
for $\tilde A_\eta$ using $\tau_{\rm iso}/\tau_0=10$ so that the time
evolution starts in the prolate phase, and
$\nu=0$ and 1. The late-time (large oblate anisotropy) behavior is
reproduced very well, with noticeable deviations at earlier times.

\subsubsection{Unstable (transverse) modes}\label{central}

With $k=0$, there are two degenerate transverse modes described by the
then coinciding equations (\ref{tij1}) and (\ref{tij2}) for the
induced currents $\tilde j^1$ and $\tilde j^2$, which
reduce to
\bea\label{tijik0}
\tilde{j}^i(\tau;\nu)&=&-\frac{m_D^2}{4}\int\frac{d\bar{y}}{(1+\frac{\tau^2\sinh^2\bar{y}}{\tau_{\rm iso}^2})^2}\Biggl\{\tilde{A}^i(\tau;\nu)-e^{i\nu\bar{\eta}_0}\tilde{A}^i(\tau_0;\nu)\nonumber\\
&&\qquad-\int_{\tau_0}^\tau d\tau' \frac{i\nu\tau e^{i\nu\bar{\eta}'}\sinh\bar{y}}{\tau_{\rm iso}^2 \sqrt{1+\frac{\tau^2\sinh^2\bar{y}}{\tau'^2}}}\tilde{A}^i(\tau';\nu)\Biggr\}+\tilde j^i_0(\tau;\nu).
\eea
This has to be solved together with
\begin{equation}
\tilde \Pi_i(\tau;\nu)=\tau \6_\tau A^i,\quad
 \frac{1}{\tau}\partial_\tau\tilde{\Pi}_i(\tau;\nu)=\tilde{j}^i(\tau;\nu)-\frac{\nu^2}{\tau^2}\tilde{A}^i(\tau;\nu).
\end{equation}

In the notation of Table~\ref{table1},
the solutions of these equations
correspond to the modes ``$-$'' and ``$\alpha$'', which become
the same at $\vartheta=0$.
This case, which contains magnetic Weibel instabilities
for oblate anisotropies, was studied already in Ref.~\cite{Romatschke:2006wg},
but with vanishing initial $W$ fields and, correspondingly, vanishing
initial currents, which now is perfectly consistent
with the Gauss law constraint $\tilde j^\tau \propto \tilde\Pi^\eta=0$.

\begin{figure}[t]
  \includegraphics[scale=1]{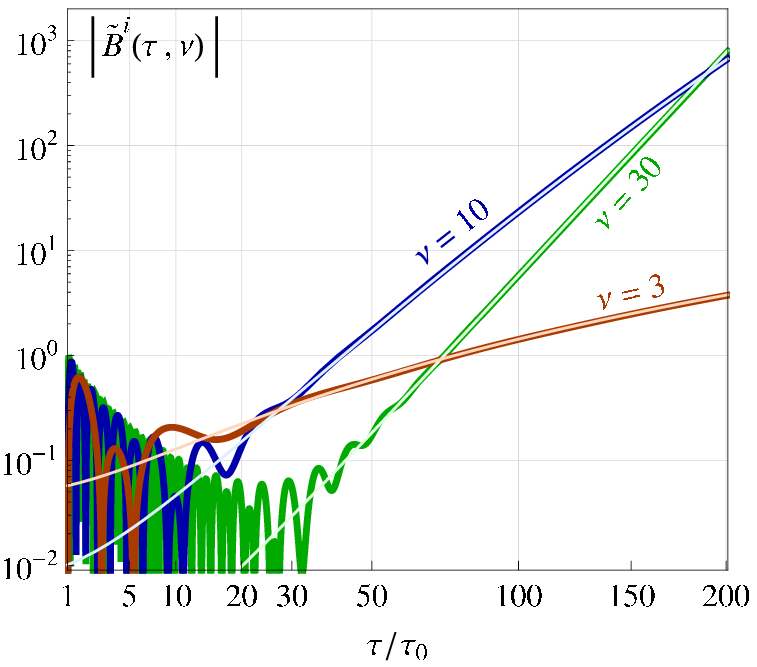}\\
\bigskip
  \includegraphics[scale=1]{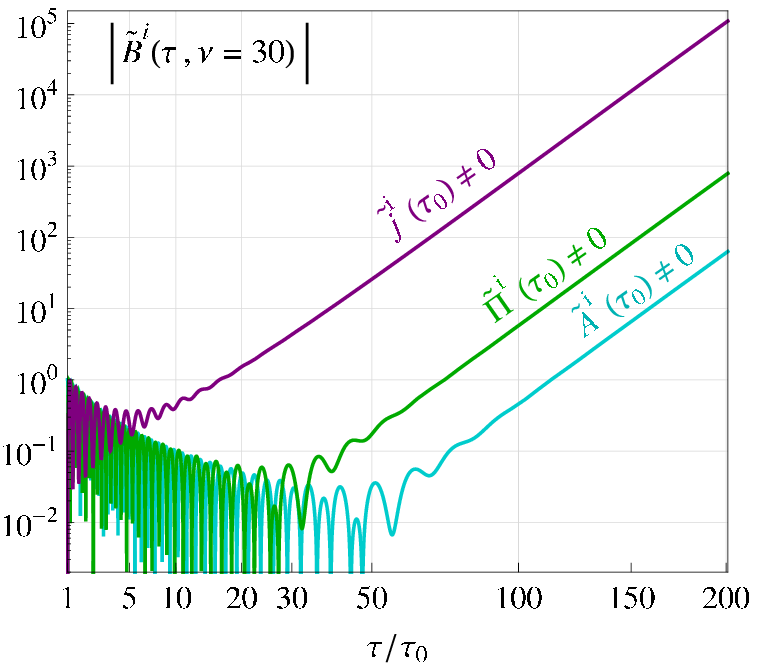}
 \caption{Magnetic (Weibel) instabilities with $k=0$ for an oblate momentum distributions from the beginning with $\tau_{\rm iso}/\tau_0=0.1$.
In the upper panel transverse magnetic fields are compared for different wave numbers $\nu$ but equal initial conditions $\tilde{\Pi}^i(\tau_0)=1$ and $\tilde{A}^i(\tau_0)=\tilde{j}^i(\tau_0)=0$. The expected analytical late time behavior, Eq.~(\ref{Aias}), is indicated by the thin light lines. In the lower plot the influence of different initial values is studied for modes with $\nu=30$.
}
 \label{fig:trans1} 
\end{figure}

In the following numerical evaluations we shall be more general
and consider separately
the initial conditions of (i) a seed electric field with only
$\tilde{\Pi}^i(\tau_0)\not=0$
, which is the case studied semi-analytically
in Ref.~\cite{Romatschke:2006wg}, (ii) a
seed magnetic field with only $\tilde{A}^i(\tau_0)\not=0$
, which was covered before
in the real-time lattice calculations of Ref.~\cite{Rebhan:2008uj},
and (iii) only $\tilde{j}^i(\tau_0)\not=0$.
(In the present linear response analysis, the most general case is
given by a linear superposition of these possibilities.)

A nonvanishing initial current $\tilde j^i_0(\tau_0)$ is provided
by any nonzero function $\tilde W_i^0(k=0,\nu;\phi,y-\eta_0)$ that is
even in its last two arguments. In the following we simply take a constant
\be
\tilde W_i^0(k=0,\nu;\phi,y-\eta_0)=C_2, 
\ee
which we found to be also
representative of some more complicated possibilities
that we have studied. Proceeding as above,
this determines the function $\tilde j^i_0(\tau)$ in 
Eqs.~(\ref{tij1}) and (\ref{tij2}) with $C_2$ proportional to
the initial values $\tilde j^i(\tau_0)$.

In the upper part of Fig.~\ref{fig:trans1} we consider the case (i) of
an initial seed electric field for $\tau_{\rm iso}/\tau_0=0.1$ and compare
with the analytic result for the late-time asymptotics (\ref{Aias}) for
various values of $\nu$. 
This is the analog to Fig.~1 of Ref.~\cite{Romatschke:2006wg} but with our larger mass 
parameter\footnote{As 
mentioned above, the recent CGC results \cite{Lappi:2007ku}
now favor this larger mass parameter, which was also considered in
Ref.~\cite{Romatschke:2006wg} but without corresponding plots.}
and less extreme initial anisotropy. 
Like Ref.~\cite{Romatschke:2006wg}
we observe a substantial
delay in the onset of plasma instabilities which is preceded by a decay
of collective fields until $\sim 10\tau_0$, suggesting an uncomfortable
suppression of Weibel instabilities by the initial strong expansion
of the plasma.

\begin{figure}
  \centering
  \includegraphics[scale=1]{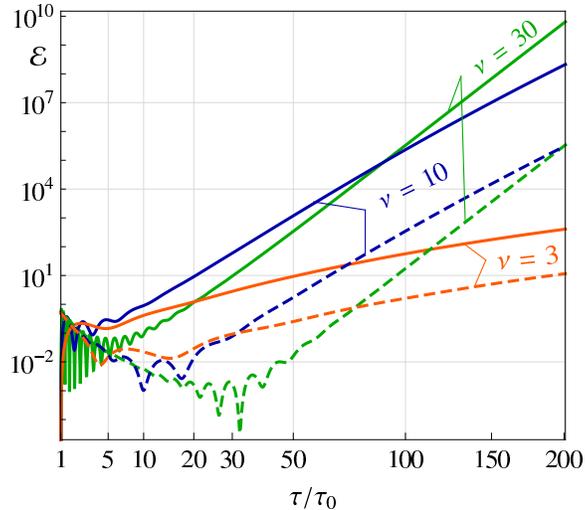}
  \caption{Total energy density of Weibel instabilities for different wave numbers $\nu$ and $k=0$, for $\tau_{\rm iso}/\tau_0=0.1$. The full lines correspond to $\tilde{j}^i(\tau_0;\nu)\neq0$, while the dashed lines are the results for $\tilde{\Pi}^i(\tau_0;\nu)\neq0$. }
  \label{fig:trans3}
\end{figure}

\begin{figure}
 \centering
 \includegraphics[scale=1]{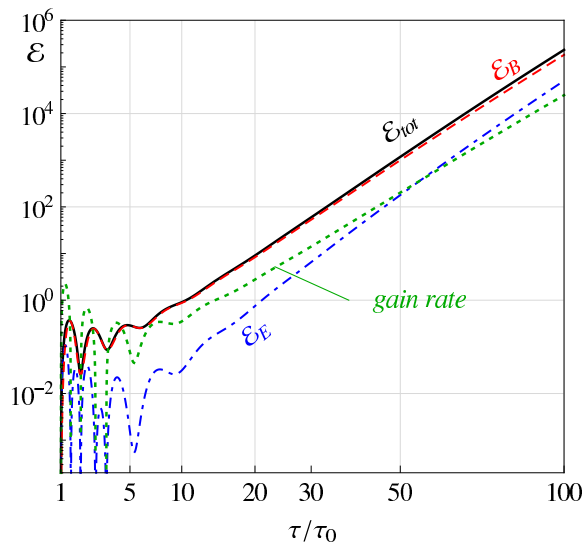}
 \caption{The total energy density $\mathcal{E}_{tot}$ for the mode with
$\nu=10$ and $\tilde{j}^i(\tau_0;\nu)\neq0$ of Fig.~\ref{fig:trans3}
and its contributions from electric ($\mathcal{E}_E$) and magnetic fields ($\mathcal{E}_B$). Additionally the gain rate (times
$\tau_0$) is shown.}
 \label{fig:gainrate}  
\end{figure}

In the lower part of Fig.~\ref{fig:trans1}, the dependence of this
behavior on initial conditions is displayed for
mode $\nu=30$. Case (ii) corresponds to
using seed magnetic fields
instead of seed electric fields, but this only 
increases the delay of plasma instabilities,
which is in line with the results of \cite{Rebhan:2008uj} where mixed
initial conditions for the fields were considered.
Surprisingly enough, with case (iii) which corresponds to 
initial fluctuations in the
currents, we find that this delay is very strongly reduced.

In Fig.~\ref{fig:trans3} the comparison between cases (i) and (iii) is
repeated for various values of $\nu$, showing now the total energy
in the collective fields, which confirms the finding of
a drastic acceleration of the onset of
plasma instabilities when there are
initial current fluctuations. In Fig.~\ref{fig:gainrate} the
energy density of one of those quickly growing modes
($\nu=10$) is decomposed into magnetic
and electric contributions together with the gain rate defined in
Eq.~(\ref{REG}).

{
From Fig.~\ref{fig:trans3} one can also easily see the
effect of having initial fluctuations in both induced currents 
and (chromo-)electromagnetic fields. Because
everything is linear, the resulting solutions are just linear
combinations, and because the onset of exponential growth
is so much quicker for the part corresponding to initial current fluctuations,
such superpositions are dominated overwhelmingly by the
latter as long as the two components have comparable
initial energy densities. Since in the physical context of heavy-ion collisions
we expect to find fluctuations in all quantities, induced currents 
($W$ fields)
as well as gauge fields, we thus consider case (iii) as being actually
representative of generic situations.
}

The strong reduction of the delay of the onset
of Weibel instabilities makes it appear much more likely that
plasma instabilities could play an important role in
the very early dynamics of a quark-gluon plasma, at least with
LHC energies, where $Q_s\sim 3$ GeV. Choosing 
$\tau_0\sim Q_s^{-1}\sim\frac1{15}$fm/c
and judging from the time it takes that the initial depletion of
energy in the fastest mode is reversed in Fig.~\ref{fig:trans3}, one may
set the scale where plasma instabilities kick in to $\sim 0.5$ fm/c,
whereas the less generic initial conditions with only seed fields
and no currents considered previously in Refs.\ \cite{Romatschke:2006wg,Rebhan:2008uj}
would have given $\sim 3$ fm/c. (RHIC energies would
give values 2--3 times higher.)

\subsection{Wave vector perpendicular to anisotropy direction}

Another comparatively simple case is provided by a wave vector which is
strictly perpendicular to the anisotropy direction, i.e.\ $\nu=0$
and $k\not=0$. In this case the integro-differential equations
for $\tilde A^1$, $\tilde A^2$ and $\tilde A_\eta$ again decouple,
but the equations for $\tilde A^1$ and $\tilde A^2$ are no longer
identical, so that we have 3 different modes. Now there are two
stable modes:
the purely electrical mode with polarization along the wave vector,
described by $\tilde A^1$, and the (``$\alpha$'') mode $\tilde A^2$,
where the electric field is 
transverse to both the wave vector and the anisotropy direction
and the magnetic field pointing in the anisotropy direction.
The third mode, $\tilde A_\eta$,
where the electric field points in the anisotropy
direction while the magnetic field is transverse 
to both the wave vector and the anisotropy direction is
stable for oblate anisotropies, but contains a magnetic Weibel
instability for the prolate case. We shall therefore now
consider $\tau_{\rm iso} > \tau_0$ so that we have an initial
period of prolate anisotropy before the expansion changes
that into an oblate one, where none of the $\nu=0$ modes is unstable.

\subsubsection{Stable modes}

\begin{figure}
 \centering
 \includegraphics[scale=1]{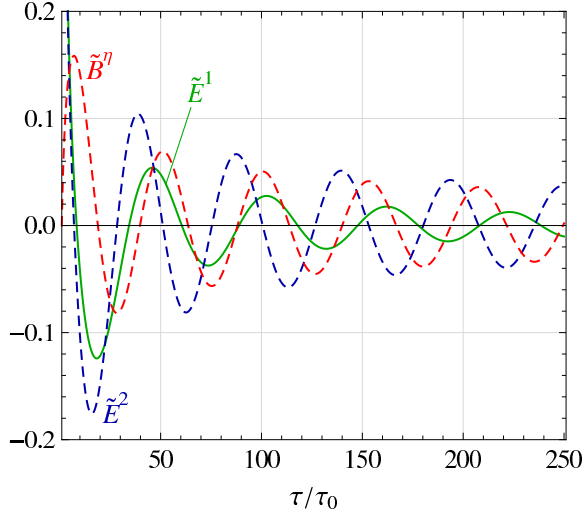}
\caption{Stable modes for wave vectors perpendicular to the anisotropy direction with $k=0.1\,\tau_0^{-1}$, initial condition $\tilde{\Pi}_i(\tau_0;k)=1$,
and $\tau_{\rm iso}/\tau_0=100$. The full line corresponds to the longitudinal plasmon mode $\tilde{E}^1$,
the dashed lines to electric and magnetic fields of transverse plasmons.
}
 \label{fig:stable1a}
\end{figure}

The purely electrical mode $\tilde A^1$
with polarization along the wave vector
again requires initial $W$ fields in order to satisfy the Gauss law
constraint. One of the simplest choices is
\be
\tilde W_1^0(k,\nu=0;\phi,y-\eta_0)=C_3 \cos\phi
\ee
with a constant $C_3$ that is proportional to
the initial value $\tilde \Pi_1(\tau_0;k)$ appearing in the
initial Gauss law constraint
\be
 \tau_0\tilde{j}^\tau(\tau_0;k)=-ik\tilde{\Pi}_1(\tau_0;k).
\ee 

Mode $\tilde A^2$, which is transverse to both wave vector and
anisotropy direction, does not need initial $W$ fields to satisfy
the Gauss law constraint. Because we are more interested in
the influence of different initial conditions on the evolution of
plasma instabilities, we shall only consider the simplest case
of $\tilde \Pi^2(\tau_0)\not=0$.

In Fig.~\ref{fig:stable1a} we compare the numerical results for the
electric fields
corresponding to the two stable modes $\tilde A^1$ and $\tilde A^2$
for $k=0.1\tau_0^{-1}$ and  $\tau_{\rm iso}/\tau_0=100$, with a
time range so that both prolate and oblate anisotropies are appearing.
As expected, only stable oscillatory behavior is found.
We observe that the purely electrical (plasmon) mode has a smaller
frequency than the transverse stable mode, which is qualitatively
similar to the familiar
behavior in the isotropic case \cite{Weldon:1982aq}. 
Perhaps more surprising is 
the rather strong attenuation during the first oscillations
which is even stronger in the longitudinal plasmon mode.

\begin{figure}[t] 
  \includegraphics[scale=1]{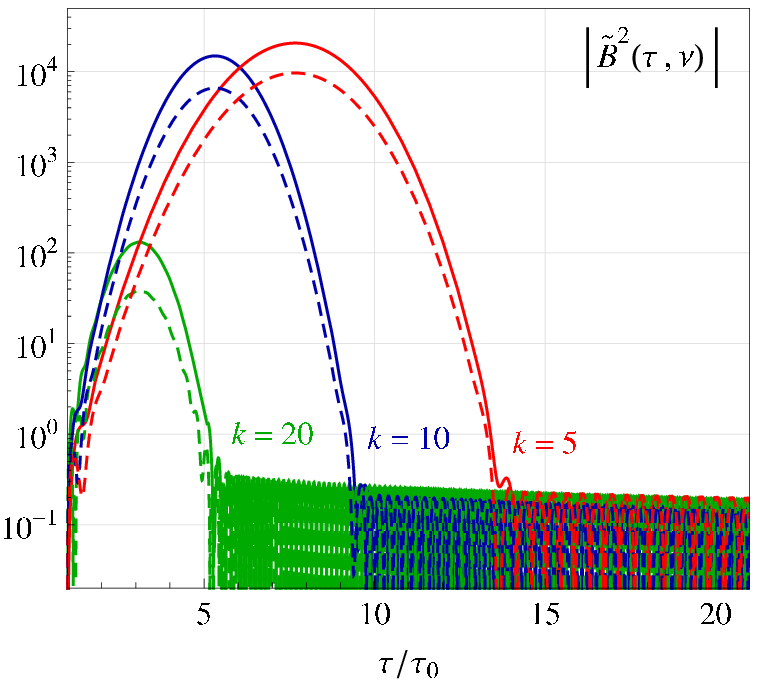}\\
\bigskip
  \includegraphics[scale=1]{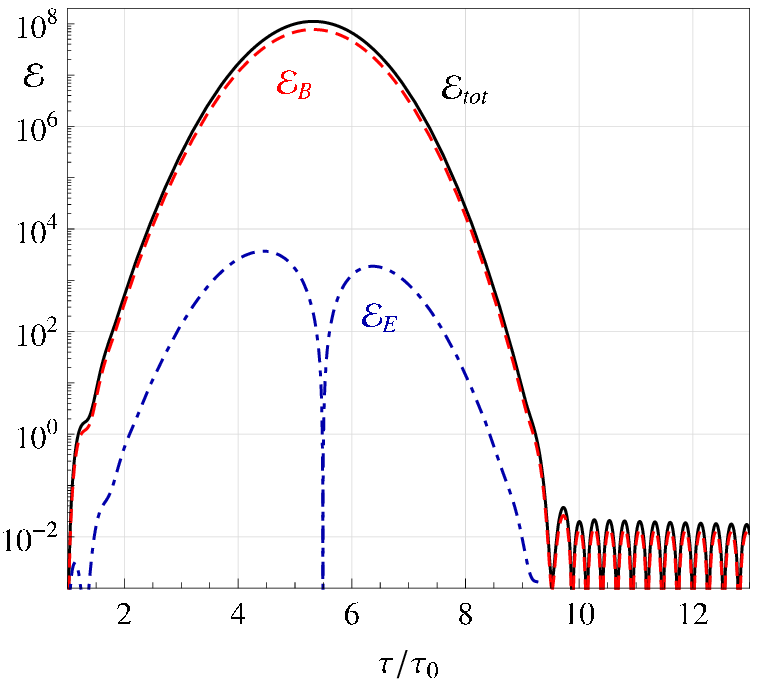}
 \caption{Magnetic Weibel instabilities for initially prolate momentum distributions with $\tau_{\rm iso}/\tau_0=10$ for different wave numbers $k$ (in units of $\tau_0^{-1}$) and two different initial conditions (upper panel). The dashed lines correspond to $\tilde{\Pi}^\eta(\tau_0;k)\neq0$ and the full lines to $\tilde{j}^\eta(\tau_0;k)\neq0$. In the lower panel the total energy density and its contributions from electric and magnetic fields are shown for the mode with $k=10\,\tau_0^{-1}$ and $\tilde{j}^\eta(\tau_0;k)\neq0$. The mass parameter in both plots has been increased to $m_D^2=1000/(\tau_{\rm iso} \tau_0)$. }
 \label{fig:prolateunstable} 
\end{figure}

\begin{figure}
 \centering
 \quad\includegraphics[scale=1]{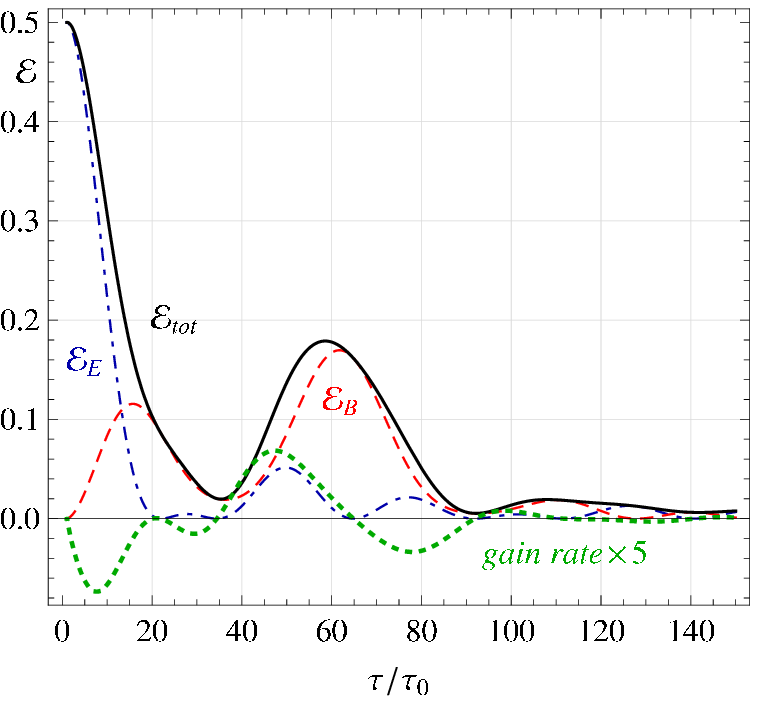}\\
\bigskip
 \includegraphics[scale=1.05]{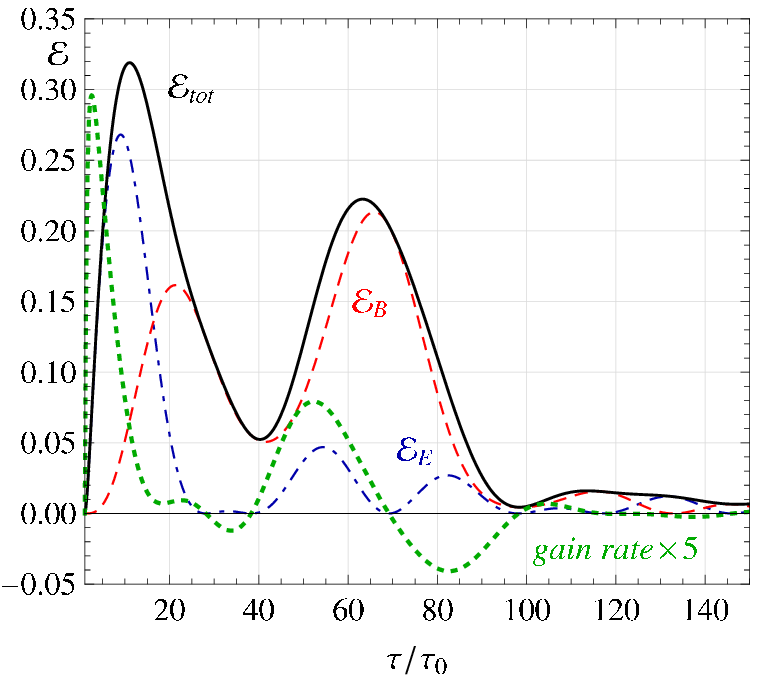}
  \caption{Total energy densities and the contributions from electric and magnetic fields of prolate-phase
Weibel instabilities with $k=0.1\tau_0^{-1}$, $\tau_{\rm iso}/\tau_0=100$, and
the CGC motivated Debye mass $m_D^2=1.285/(\tau_{\rm iso}\tau_0)$.
The upper plot corresponds to the initial condition $\tilde \Pi^\eta(\tau_0)=1$, the lower one to $\tilde j^\eta(\tau_0)\not=0$. The gain rate 
(times $\tau_0$, dotted line)
is increased by a factor 5 to make it better visible.}
 \label{fig:unstable2}
\end{figure}

\subsubsection{Weibel instability during prolate phase}

In the time evolution of $\tilde A_\eta(\tau;k,\nu=0)$ we expect
to find magnetic Weibel instabilities for prolate anisotropies,
thus only as long as $\tau<\tau_{\rm iso}$. In contrast to Weibel instabilities
for oblate anisotropies there is now only one unstable mode, namely
with electric field transverse to the wave vector and pointing along
the direction of anisotropy, and the magnetic field transverse to both.

We again consider initial conditions with and without initial currents.
Nonzero initial induced currents $\tilde j^\eta$ can be set up most simply by
choosing a constant
$\tilde W_\eta^0(k,\nu=0;\phi,y-\eta_0)=C_4$, which we compare
with vanishing $W_\eta^0$ and nonzero initial electric field,
$\tilde \Pi^\eta(\tau_0)\not=0$.

As before we choose $\tau_{\rm iso}\gg\tau_0$ to have an extended phase
where the free-streaming plasma has prolate anisotropy, and we again
stop the numerics only after the momentum distribution has become oblate.

Unfortunately the numerical solutions show rather little activity with
the mass parameter $m_D$ extracted from CGC calculations. We therefore
choose the much higher value $m_D^2=1000/(\tau_0\tau_{\rm iso})$ at first and
consider the situation for our standard value only thereafter.
Numerical results for the higher mass parameter and
$\tau_{\rm iso}=10\,\tau_0$ are displayed in 
Fig.~\ref{fig:prolateunstable}a, which exhibit a pronounced instability
that shuts off when the degree of prolate anisotropy becomes too small
for a given value of $k$. After that point in time the mode
decays for about as long as it was growing initially,
ending in stable oscillations. Dashed lines correspond to nonzero
initial values of $\tilde \Pi^\eta(\tau_0)$ and full lines to nonzero initial
currents $\tilde j^\eta\not=0$, normalized such that the two different
initial conditions have equal amplitude in the final oscillations.
Since the solutions with nonzero initial currents reach larger maximal
values we again find, although to a lesser degree, that such initial conditions
are more efficient seeds for unstable modes.

In Fig.~\ref{fig:prolateunstable}b the energy content in magnetic and
electric fields is shown for one of the unstable modes, demonstrating that
the energy is predominantly in magnetic fields, as expected for a Weibel
instability. Near the point where the instability stops the electric
field changes sign. 

In Fig.~\ref{fig:unstable2} we finally consider our much smaller
standard choice $m_D^2\tau_0\tau_{\rm iso}=1.285$ and again the two
initial conditions of nonzero initial electric field (upper plot,
where $\tilde \Pi^\eta(\tau_0)=1$) and
nonzero initial current (lower plot), normalized so that
the final oscillations have equal amplitude to first case. 
Also shown is the gain rate,
defined in Eq.~(\ref{REG}), times $\tau_0$.  Notice that the plot is
now linear instead of logarithmic.  In order to observe some
instability, we need to consider much stronger initial anisotropy and
$k\ll\tau_0^{-1}$. Fig.~\ref{fig:unstable2} shows the various energy
components for $\tau_{\rm iso}=100\tau_0$ and $k=0.1\tau_0^{-1}$.

Since the instability is now a magnetic one, its presence is best
judged from the magnetic energy content. With initial electric seed
field, the dominant effect is the transfer of the electric field
energy to the hard particle background. The magnetic field energy
does increase, albeit non-monotonically, without reaching the
initial energy density supplied by the seed field.
In the case of nonzero initial currents, there is a sharp initial
increase in the energy density, which is however predominantly
electric. The magnetic energy density eventually
increases, too, again non-monotonically, and only slightly higher than
in the case with seed electric field.

We thus find that for our CGC-inspired mass parameter, the Weibel
instabilities in the prolate phase (where the amount of
prolate anisotropy decreases rapidly) are rather weak
when compared with the required energy densities in the
initial seed configuration.

\subsection{General wave vectors and electric instabilities}

Up to now we have only considered the special situations where the
integro-differential equations for $\tilde{A}^1$ and $\tilde{A}^\eta$ were
decoupled. We now turn to the more general case where this is no longer
the case because both $k$ and $\nu$ are non-zero. Such Fourier components
correspond to a physical wave vector whose angle with the
$\eta$ (or $z$) axis increases with time according to $\vartheta=\arctan(\tau k/\nu)$.

In the oblate anisotropic case the growth rate of unstable modes 
decreases with increasing $\vartheta$. For the $\alpha$ mode the rate
tends to zero as $\vartheta\to\pi/2$ and for the ``$-$'' mode already
at $\le\pi/4$. It is therefore of some interest to future
3+1-dimensional simulations of non-Abelian plasma instabilities to determine
the range of wave numbers $k$ for which 
the corresponding modes play an important
part in the evolution. 

Another interesting aspect of the generic case is that it allows us to
study electric instabilities which appear in the ``$-$'' mode.

\subsubsection{Electric instabilities for oblate anisotropy}

When solving the coupled integro-differential equations for $\tilde{A}^1$ and $\tilde{A}^\eta$ for nonzero $k$ and $\nu$ we have again to take care of the nontrivial
Gauss law constraint (\ref{GLconstr}) at $\tau_0$. To do so, we adopt the initial
data for $W$ field of Eq.~(\ref{eq:W_eta1}) also for nonzero $k$,
\be\label{Wetageneral}
\tilde W_\eta^0(k,\nu;\phi,y-\eta_0)=C_5 \tanh(y-\eta_0),
\ee
where the constant $C_5$ is proportional to
the initial electric field component parallel to the wave vector,
\be
E_\parallel=\frac{k\tilde{\Pi}^1+\nu\tilde{\Pi}^\eta}{\sqrt{k^2\tau^2+\nu^2}}.
\ee
The electric field, lying in the 1-$\eta$ plane, has in general also
a component transverse to the wave vector, which we shall denote $E_\perp$ in
this subsection. (The magnetic field is of course purely transverse and
points in the 2-direction.)

An initial $W$ field given by (\ref{Wetageneral}) gives rise to an
initial charge density $\tilde j_0^\tau(\tau_0)$ as required
by the Gauss law constraint, but zero initial spatial
currents. In order to also have nonzero initial current components $\tilde
j^\eta$ and $\tilde j^1$, we additionally add the components
\be
\tilde W_{\eta,1}^0(k,\nu;\phi,y-\eta_0)=C_{6,7}.
\ee

As discussed in Sect.~\ref{sect:modes}, we can expect to find
an electric instability for wave vectors with nonvanishing
$\vartheta<\pi/4$. Since $\vartheta$ increases with time, we
choose small initial values of $\vartheta$, which over some time
satisfy the criterion of being within a $45^\circ$ cone about the $\eta$ axis.

\begin{figure}
 \centering
 \includegraphics[scale=1]{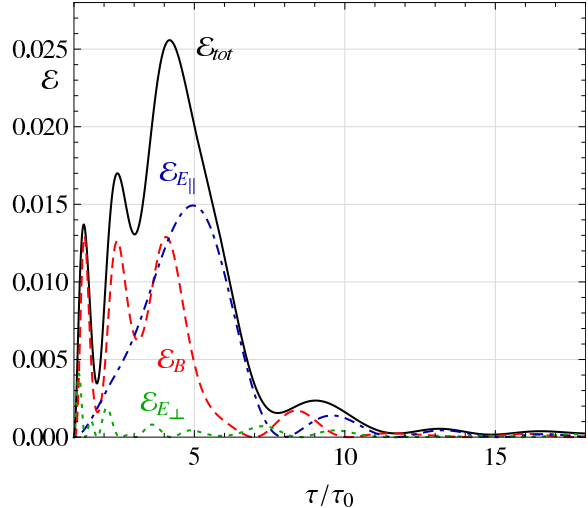}
 \caption{Total energy density and the contributions from electric and magnetic fields for a mixed $\tilde A^{1,\eta}$ mode with $k\tau_0=1$ and $\nu=10$
exhibiting an electric instability.
The electric energy density is separated into a (dominant) longitudinal and a (small) transverse part with respect to the wave vector. The remaining parameters are $m_D^2=10/(\tau_{\rm iso}\tau_0)$ and $\tau_{\rm iso}/\tau_0=0.1$; the
initial condition is $\tilde{j}^1(\tau_0;k,\nu)\neq0$.}
 \label{fig:electric_instability}
\end{figure}

In Fig.\ \ref{fig:electric_instability} the time evolution of the energy density for a mode with nonzero initial $\tilde j^1$ and $k\tau_0=1$, $\nu=10$ and is shown, for which we can expect an electric instability only for times smaller than $10\tau_0$.
Again we consider a larger mass parameter $m_D^2=10/(\tau_{\rm iso}\tau_0)$ to find more significant results and indeed we notice that the total energy density rises initially 
with an significant electric component that is almost entirely longitudinal. After the maximum at about 4$\tau_0$ we observe a strong decay and only small plasma oscillations
after $10\tau_0$.

\subsubsection{General magnetic instabilities for oblate anisotropy}

For general wave vector and oblate anisotropy, 
the purely magnetic instabilities reside in the
$\tilde A^2$ modes for all $\vartheta<\pi/2$, but with vanishing
growth rates as $\vartheta\to\pi/2$. Nonvanishing initial currents
can be simply taken into account by choosing a constant 
$\tilde W_2^0(k,\nu;\phi,y-\eta_0)=C_8$, with $C_8$ proportional to
$\tilde j^2(\tau_0)$.

\begin{figure}
 \centering
 \includegraphics[scale=1]{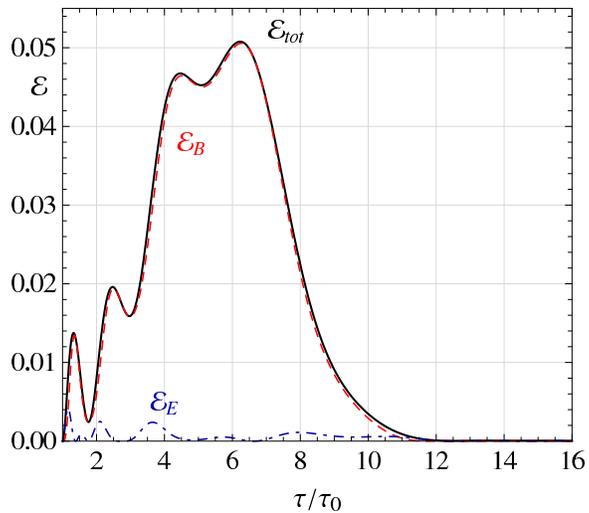}
  \caption{Total energy density and its electric and magnetic contributions for the purely transverse $\tilde A^2$ mode with parameters as in 
Fig.~\ref{fig:electric_instability} and
initial condition $\tilde{j}^2(\tau_0;k,\nu)\neq0$.}
 \label{fig:mix2}
\end{figure}

Using the same parameters as in Fig.\ \ref{fig:electric_instability},
but now with nonzero initial $\tilde j^2$,
we find an instability that is operative up to the somewhat larger
time of about 6.5 $\tau_0$ (see Fig.\ \ref{fig:mix2}) and which is
almost completely in magnetic fields. For this set of parameters it was 
in fact crucial to
have nonzero initial currents---with vanishing initial currents and only
initial gauge fields we only found decreasing solutions.

\begin{figure}
\centering
\includegraphics[scale=1]{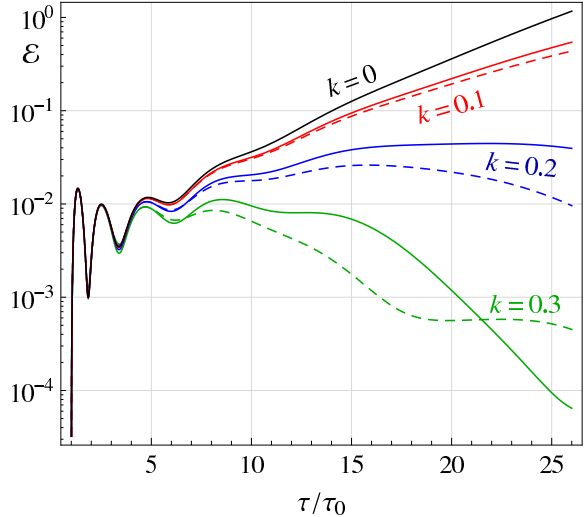}
\caption{Time evolution of unstable modes with $\nu=10$ and
various values of $k$ (in units of $\tau_0^{-1}$) and 
$\tau_{\rm iso}/\tau_0=0.1$. Full lines correspond to the
purely transverse $\tilde A^2$ modes,
dashed lines to the mixed $\tilde A^{1,\eta}$ modes,
both with initial conditions of nonzero initial currents.}
\label{fig:mix}
\end{figure}

Returning to our standard choice of mass parameter, Fig.~\ref{fig:mix}
displays the time evolution of the unstable modes 
with different values of $k$, the full lines corresponding to the
$\tilde A^2$ modes, and the dashed ones to mixed $\tilde A^{1,\eta}$ ones.
This shows that the most efficient plasma instabilities
in the phase of oblate anisotropies are concentrated
in the range $k \lesssim 0.2\,\tau_0^{-1}$.

\section{Summary and Conclusions}

In this paper we have generalized the semi-analytical analysis
of plasma instabilities in an anisotropically expanding plasma of
Ref.~\cite{Romatschke:2006wg} to general orientations of wave vectors
and all possible polarizations of the individual Fourier modes.
Moreover we have generalized to arbitrary initial data
in both the collective gauge fields and the $W$ fields of the
hard loop formalism, which correspond to colored fluctuations in the
hard particle distribution and thus directly to the induced currents.

Besides the well-studied magnetic (Weibel) instabilities of a plasma with
oblate anisotropy, we have also considered plasma instabilities
involving growing electric fields parallel to the wave vector.
For the latter, the wave vector needs to have a nonzero angle with
respect to the axis of expansion, which then increases with time,
eventually shutting off such instabilities in the expanding case.
We have also considered a plasma that starts
with prolate anisotropy which after some time
turns into an oblate one. Such instabilities have
occasionally been conjectured to be the most interesting
for isotropization in heavy-ion collisions \cite{Randrup:2003cw}.
However we found that this type of Weibel instabilities requires
plasma densities much larger than those suggested by
CGC calculations.

For Weibel instabilities in the oblate phase, Ref.~\cite{Romatschke:2006wg}
has previously observed an uncomfortably long delay before they
overcome the depletion of the energy in initial fields
due to the (free-streaming) expansion of the plasma.
With parameters taken over from CGC calculations, there
seemed to be very little room for plasma instabilities
for the available energies and plasma lifetimes at RHIC, while
energies and plasma lifetimes expected for heavy-ion collisions
at the LHC would
make an important role conceivable, if the quark-gluon matter
to be produced there turns out to be sufficiently weakly
coupled to behave as a plasma. 

{
However, Ref.~\cite{Romatschke:2006wg} has considered only
initial fluctuations in collective fields as seeds for
plasma instabilities, while physically one should expect
fluctuations in both collective fields and in the
initial hard particle distribution.
With our more general initial conditions that allow also
for fluctuations in the initial hard particle distribution 
and the
corresponding induced currents, we find much more favorable
conditions for plasma instabilities.
When the initial fluctuations in (only) induced currents are such that
they give rise to the same energy content in collective modes as
considered in the case of only initial field fluctuations, 
a surprisingly stark reduction of the delays
of plasma instabilities by almost an order of magnitude was obtained.
Because of linearity in the weak field regime,
this implies that
the generic case of fluctuations of comparable
strength in both induced currents and collective fields
is overwhelmingly dominated by the modes corresponding to only
initial current fluctuations.
}

In Sect.~\ref{central} we have concluded that for LHC energies
the time scales for
plasma instabilities to set in are of the order of $\sim 0.5$ fm/c
when initial current fluctuations
are considered, while for RHIC energies these values would be
about 2--3 times larger. Although for a significant backreaction of the plasma
instabilities on the anisotropic hard particle distribution
one would presumably have to consider times that are
somewhat larger, this still seems to keep this mechanism very interesting
at least for LHC energies.

Finally, we should emphasize that the present analysis was carried
out in the weak-field (linear-response) regime. 
{
The nonlinear regime of quark-gluon plasma instabilities in
the case of boost-invariant expansion was considered for an
effectively 1+1-dimensional evolution in Ref.\ \cite{Rebhan:2008uj}.
These results remain valid as far as the specific non-Abelian dynamics
is concerned, but the uncomfortably long delay of the onset of
the instabilities largely disappears by considering also initial fluctuations
in the induced currents (equivalently, in the $W$ fields).
}
However, ultrarelativistic
plasma instabilities have been found to behave very differently
in the regime of nonperturbatively large non-Abelian gauge field
when a full 3+1-dimensional situation is considered.
Work in this direction is in progress, for which the present
semi-analytical results will provide important cross-checks
for the initial stages of the evolution of non-Abelian plasma instabilities.

\section*{Acknowledgements} 

We thank Paul Romatschke and Mike Strickland for useful discussions,
and Andreas Ipp for a careful reading of the manuscript.
This work was supported by the Austrian science foundation FWF,
 project no.\ P19526. 

\appendix
\section{Analytical late-time behavior}
\label{app}

For modes with wave vector parallel to the anisotropy direction,
i.e.\ $k=0$ and $\nu\neq0$, which is the case studied before in
Ref.~\cite{Romatschke:2006wg}, the expressions for the induced currents
simplify, and
it is possible to study 
the late-time behavior of single modes analytically by expanding 
the contributions to the memory integrals around $\tau'=\tau$.
Late-time behavior in our free-streaming approximation means
extreme anisotropy, characterized by $\tau_{\rm iso}/\tau
\equiv \theta \ll 1$. 
In this appendix we recapitulate the analytical results
of Ref.~\cite{Romatschke:2006wg}, filling in some details
and also show that the late-time behavior is not modified
by the necessity of including initial values for the longitudinal
current in the case of longitudinal modes.

\subsection{Transverse modes}\label{AppT}
With $k=0$, 
the induced currents $\tilde{j}^i(\tau;\nu)$ are given by Eq.~(\ref{tijik0}),
where for simplicity we set $\tilde A^i(\tau_0;\nu)=0$ and $\tilde j^i_0(\tau;\nu)=0$. Expanding the integrand of the memory integral around $\tau'=\tau$ yields
\bea
\frac{i\nu\tau e^{i\nu\bar{\eta}'}\sinh\bar{y}}{\tau_{\rm iso}^2 \sqrt{1+\frac{\tau^2\sinh^2\bar{y}}{\tau'^2}}}&=&\frac{i\nu\tau}{\tau_{\rm iso}^2}\Biggl(\tanh\bar{y}+i\nu\tanh^2\bar{y}\Bigl(1-\frac{\tau}{\tau'}\Bigr)\nonumber\\
&&\qquad+\tanh^3\bar{y}\Bigl(1-\frac{\tau}{\tau'}\Bigr)+O\Bigl[\Bigl(1-\frac{\tau}{\tau'}\Bigr)^2\Bigr]\Biggr).
\label{eq:expansion}
\eea
Terms odd in $\bar{y}$ give no contribution and neglecting the higher orders we
obtain
\bea
\tilde{j}^i(\tau;\nu)&\simeq&-\frac{m_D^2}{4}\int\frac{d\bar{y}}{(1+\frac{\tau^2\sinh^2\bar{y}}{\tau_{\rm iso}^2})^2}\Biggl\{\tilde{A}^i(\tau;\nu)\nonumber\\
&&\qquad+\int_{\tau_0}^\tau d\tau' \frac{\nu^2\tau \tanh^2\bar{y} }{\tau_{\rm iso}^2}\Bigl(1-\frac{\tau}{\tau'}\Bigr)\tilde{A}^i(\tau';\nu)\Biggr\}.
\label{eq:jlate}
\eea

Using
\bea
\int\frac{d\bar{y}}{(1+{\theta^{-2}\sinh^2\bar{y}})^2}&=&\frac{(\theta^{-2}-2)\arctan(\sqrt{\theta^{-2}-1})}{(\theta^{-2}-1)^{3/2}}+\frac{1}{\theta^{-2}-1}\nn\\&=&\frac{\pi \theta}{2}-\frac{\pi \theta^3}{4}+O(\theta^4)
\eea
and
\bea
\int\frac{d\bar{y}\tanh^2\bar{y}}{(1+\theta^{-2}\sinh^2\bar{y})^2}&=&\frac{(2+\theta^{-2})\arctan(\sqrt{\theta^{-2}-1})}{(\theta^{-2}-1)^{5/2}}-\frac{3}{(\theta^{-2}-1)^2}\nonumber\\
&=&\frac{\pi \theta^3}{2}+O(\theta^4)
\eea
for $\theta\equiv \tau_{\rm iso}/\tau\ll 1$,
the transverse current reduces to
\begin{equation}
\tilde{j}^i(\tau;\nu)\simeq-\frac{\mu}{\tau}\tilde{A}^i(\tau;\nu)-\frac{\mu\nu^2}{\tau^2}\int_{\tau_0}^\tau d\tau'\tilde{A}^i(\tau';\nu)\Bigl(1-\frac{\tau}{\tau'}\Bigr),
\end{equation}
where $\mu=m_D^2\pi \tau_{\rm iso}/8$ and only terms up to linear order in $\tau_{\rm iso}/\tau$ have been kept. The equation of motion for the transverse gauge fields are 
\begin{equation}
\Bigl(\frac{1}{\tau}\partial_\tau \tau \partial_\tau +\frac{\nu^2}{\tau^2}\Bigr)\tilde{A}^i(\tau;\nu)=\tilde{j}^i(\tau;\nu)
\end{equation}
and acting with $\partial_\tau^2\tau^2$ on it we eventually obtain an ordinary differential equation for each mode $\nu$
\begin{equation}
\Bigl(\partial_\tau^2\tau\partial_\tau\tau\partial_\tau+\nu^2\partial_\tau^2+\mu\partial_\tau^2\tau-\frac{\mu\nu^2}{\tau}\Bigr)\tilde{A}^i(\tau;\nu)\simeq0.
\end{equation}
Simple results for the gauge fields are only obtained for very infrared modes $\nu\ll1$, where all terms proportional to $\nu^2$ can be neglected, or for high momentum modes $\nu\gg1$, where only those terms proportional to $\nu^2$ contribute. We find
\begin{align}
\tilde{A}^i(\tau;\nu\ll1)\simeq c_1J_0(2\sqrt{\mu\tau})+c_2Y_0(2\sqrt{\mu\tau}),
\end{align}
which is a stable oscillatory solution ($J_n(x)$ and $Y_n(x)$ are Bessel functions of the first and second kind, respectively), and
\begin{align}
\tilde{A}^i(\tau;\nu\gg1)\simeq c_1\sqrt{\tau}I_1(2\sqrt{\mu\tau})+c_2\sqrt{\tau}K_1(2\sqrt{\mu\tau}),
\end{align}
with $c_{1,2}$ being constants. The modified Bessel functions $K_n$ and $I_n$ have the asymptotic behavior $K_n(x)\simeq\exp(-x)/\sqrt{2\pi x}$ and $I_n(x)\simeq\exp(x)/\sqrt{2\pi x}$, where the latter describes a rapidly growing mode. Therefore we expect that large $\nu$ modes will be dominant at sufficiently late times with a behavior of
\begin{equation}
\tilde{A}^i(\tau)\sim\tau^{1/4}\exp(2\sqrt{\mu\tau}).
\end{equation}
For $\nu\sim1$ the solutions can be written in terms of generalized hypergeometric functions ${}_2F_3$ and a Meijer $G$ function \cite{Romatschke:2006wg}. The dominant contribution is
\begin{equation}\label{Aias}
\tilde{A}^i(\tau;\nu)/\tau\sim {}_2F_3\Bigl(\frac{3-\sqrt{1+4\nu^2}}{2},\frac{3+\sqrt{1+4\nu^2}}{2};2,2-i\nu,2+i\nu,-\mu\tau\Bigr),
\end{equation}
which is compared with the full semi-analytical result in Fig.\ \ref{fig:trans1}.

\subsection{Longitudinal modes}\label{AppL}

\begin{figure}
\centering
\includegraphics[scale=1]{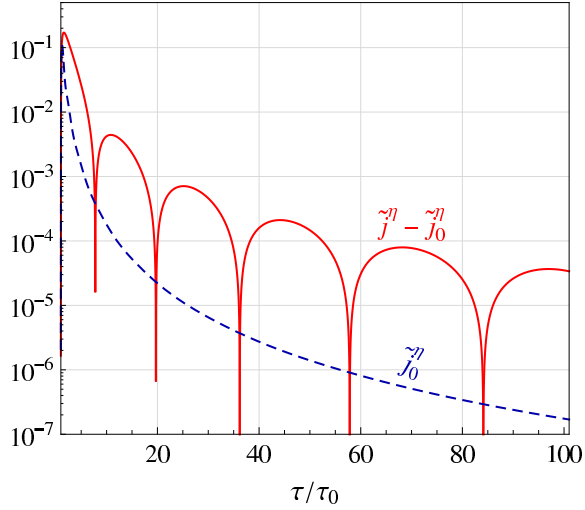}
\caption{The contribution from to the current proportional to the initial conjugate momentum $\tilde{j}^\eta_0$ is negligible compared to the rest at late times. This data is for $\nu=10$ and $\tau_{\rm iso}/\tau_0=0.01$.}
\label{fig:j0}
\end{figure}

For longitudinally polarized gauge fields we proceed analogously. 
In this case it is not a priori admissible to drop the
the term $\tilde{j}^\eta_0(\tau;\nu)$ term in Eq.~(\ref{tijeta})
because of the Gauss law constraint.
However,
numerically, we notice that at late times this contribution is negligible compared to the rest of the current, as can be seen in figure \ref{fig:j0}. 

Omitting both $\tilde{j}^\eta_0(\tau;\nu)$ and the term
proportional to $\tilde A_\eta(\tau_0;\nu)$, we can approximate 
Eq.~(\ref{tijetak0}) by
\bea
\tilde{j}^\eta(\tau;\nu)&\simeq&\frac{m_D^2}{2\tau_{\rm iso}^2}\int\frac{d\bar{y}\sinh^2\bar{y}}{(1+\frac{\tau^2\sinh^2\bar{y}}{\tau_{\rm iso}^2})^2}\Biggl\{\tilde{A}_\eta(\tau;\nu)\nonumber\\
&&\qquad+i\nu\tanh^2\bar{y}\int_{\tau_0}^\tau d\tau'\tilde{A}_\eta(\tau';\nu)\frac{1}{\tau'^2}\Bigl(1-\frac{\tau}{\tau'}\Bigr)\Biggr\},
\eea
where we have used again (\ref{eq:expansion}), but with the factor $1/\tau_{\rm iso}^2$ replaced by $1/\tau'^2$.
Using
\bea\label{Nanal}
\int\frac{d\bar{y}\,\sinh^2\bar{y}}{(1+\theta^{-2}\sinh^2\bar{y})^2}&=&\frac{\arctan(\sqrt{\theta^{-2}-1})}{(\theta^{-2}-1)^{3/2}}+\frac{1}{\theta^{-2}-\theta^{-4}}\nn\\&=&\frac{\pi \theta^3}{2}+O(\theta^4)
\eea
and
\begin{equation}
\int\frac{d\bar{y}\,\sinh^2\bar{y}\tanh^2\bar{y}}{(1+\theta^{-2}\sinh^2\bar{y})^2}=2\theta^4+O(\theta^5)
\end{equation}
for $\theta\equiv \tau_{\rm iso}/\tau\ll 1$, we obtain
\begin{align}
\tilde{j}^\eta(\tau;\nu)\simeq\frac{2\mu}{\tau^3}\tilde{A}_\eta(\tau;\nu)+\frac{8\mu\nu^2\tau_{\rm iso}}{\pi\tau^3}\int_{\tau_0}^\tau d\tau' \tilde{A}_\eta(\tau';\nu)\frac{1}{\tau'^2}\Bigl(1-\frac{\tau}{\tau'}\Bigr).
\end{align}
By acting with $\partial_\tau^2\tau^2$ on this expression we find
\begin{align}
\partial_\tau^2\big(\tau^2\tilde{j}^\eta(\tau;\nu)\bigr)\simeq 2\mu\partial_\tau^2\Big(\frac{\tilde{A}_\eta(\tau;\nu)}{\tau}\Bigr)-\frac{8\mu\nu^2\tau_{\rm iso}}{\pi\tau^4}\tilde{A}_\eta(\tau;\nu),
\end{align}
where we can neglect the second part for very large $\tau$. The equation of motion for the longitudinal gauge fields therefore becomes
\begin{equation}
\Bigl(\partial_\tau \frac{1}{\tau}\partial_\tau+\frac{2\mu}{\tau^2}\Bigr)\tilde{A}_\eta(\tau;\nu)\simeq0
\end{equation}
and the late-time behavior is given by
\begin{equation}\label{Aetaas}
\frac{\tilde{A}_\eta(\tau;\nu)}{\tau}\simeq c_1J_2(2\sqrt{2\mu\tau})+c_2Y_2(2\sqrt{2\mu\tau}).
\end{equation}
This corresponds to stable and oscillatory solutions.

\bibliographystyle{prsty}
\bibliography{ar,tft,qft}
\end{document}